\documentclass[acmsmall,screen]{acmart}

\AtBeginDocument{%
  \providecommand\BibTeX{{%
    \normalfont B\kern-0.5em{\scshape i\kern-0.25em b}\kern-0.8em\TeX}}}

\setcopyright{acmcopyright}
\copyrightyear{2023}
\acmYear{2023}
\acmDOI{10.1145/3571157}

\acmJournal{CSUR}
\acmVolume{55}
\acmNumber{12}
\acmArticle{244}
\acmMonth{3}

\usepackage{graphicx}
\usepackage{caption}
\graphicspath{{./figure/}}
\usepackage{soul}
\usepackage{color}
\usepackage{enumerate}
\usepackage{multirow}
\usepackage{amsmath}

\usepackage{amssymb}
\usepackage[linesnumbered,ruled]{algorithm2e}
\usepackage{booktabs}
\usepackage{array}
\usepackage{subfig}
\usepackage{url}
\usepackage{tabularx}
\usepackage{bm}
\usepackage{enumitem}
\usepackage{pifont}
\newcommand{\cmark}{\ding{51}}%
\newcommand{\xmark}{\ding{55}}%
\newcommand{\matr}[1]{\bm{#1}}
\newcommand{\vect}[1]{\bm{#1}}
\usepackage{xcolor}
\usepackage{textcomp}
\usepackage[commandnameprefix=ifneeded,final]{changes}
\usepackage{makecell}

\begin{document}

\title{A Systematic Survey of General Sparse Matrix-Matrix Multiplication}

\author{Jianhua Gao}
\email{gjh@bit.edu.cn}
\author{Weixing Ji}
\email{jwx@bit.edu.cn}
\authornote{Weixing Ji is the corresponding author.}
\author{Fangli Chang}
\email{cfl@bit.edu.cn}
\author{Shiyu Han}
\email{bit_hsy@bit.edu.cn}
\author{Bingxin Wei}
\email{bit_wbx@bit.edu.cn}
\author{Zeming Liu}
\email{sakusho@bit.edu.cn}
\author{Yizhuo Wang}
\email{frankwyz@bit.edu.cn}
\affiliation{%
  \department{School of Computer Science and Technology}
  \institution{Beijing Institute of Technology}
  \streetaddress{No. 5, South Street, Zhongguancun, Haidian District}
  \city{Beijing}
  \postcode{100081}
  \country{China}
} 

\renewcommand{\shortauthors}{J. Gao et al.}

\begin{abstract}
  General Sparse Matrix-Matrix Multiplication (SpGEMM) has attracted much attention from researchers in graph analyzing, scientific computing, and deep learning. Many optimization techniques have been developed for different applications and computing architectures over the past decades. The objective of this article is to provide a structured and comprehensive overview of the researches on SpGEMM. Existing researches have been grouped into different categories based on target architectures and design choices. Covered topics include typical applications, compression formats, general formulations, key problems and techniques, architecture-oriented optimizations, and programming models. The rationales of different algorithms are analyzed and summarized. This survey sufficiently reveals the latest progress of SpGEMM research to 2021. Moreover, a thorough performance comparison of existing implementations is presented. Based on our findings, we highlight future research directions\replaced{, which encourage better design and implementations in later studies.}{. Later studies can leverage our findings to encourage better designs and implementations.}
\end{abstract}



\maketitle

\section{Introduction}\label{sec:Introduction}
SpGEMM is a special case of general matrix multiplication (GEMM) when two input matrices are sparse matrices. It is a fundamental and expensive computational kernel in numerous scientific computing applications and graph algorithms, such as algebraic multigrid solvers \cite{BellDO12}\cite{Ballard2016}, triangle counting \cite{Davis2018}\cite{Cohen2009}\cite{Wolf2017}\cite{Azad2015}, multi-source breadth-first searching \cite{Gilbert2006}\cite{Then2014}\cite{BulucG11}, the shortest path finding \cite{Chan2007}, colored intersecting \cite{Kaplan2006}\cite{Deveci2016parallel}, and subgraphs matching \cite{Virginia2006}\cite{Buluc2011}. Hence, the optimization of SpGEMM has the potential to impact a wide variety of applications.

To the best of our knowledge, this is the first survey paper that overviews the developments of SpGEMM over the past decades. The goal of this survey is to present a working knowledge of the underlying theory and practice of SpGEMM for solving large-scale scientific problems, and provide an overview of the algorithms, data structures, and libraries available. This survey covers the sparse formats, application domains, challenging problems, architecture-oriented optimization techniques, and performance evaluation of available implementations. This study follows the guidelines of the systematic literature review proposed by Kitchenham \cite{Kitchenham2004}, which was initially used in medical science but later gained interest in other fields as well. According to the three main phases: planning, conducting, and reporting, we have formulated the following research questions:
    
\textbf{RQ1}: What are the applications of SpGEMM, and how are they formulated?

\textbf{RQ2}: What is the current status of SpGEMM research?
    
\textbf{RQ3}: How do the state-of-the-art SpGEMM implementations perform? 
    
\textbf{RQ4}: What challenges could be inferred from the current research effort?

Regarding RQ1, this study presents a detailed introduction to three typical applications and addresses how SpGEMM is used in these applications. This will give an insight into the requirements of SpGEMM in solving real problems. In RQ2, the study looks at existing techniques that were proposed in recent decades from different angles. Regarding RQ3, we perform some performance evaluations to have a general idea about the performance of these implementations on prevailing hardware platforms. Finally, in RQ4, we summarize the challenges and future research directions according to our investigative results.

\begin{figure}[!htbp]
\centering
\includegraphics[width=4in]{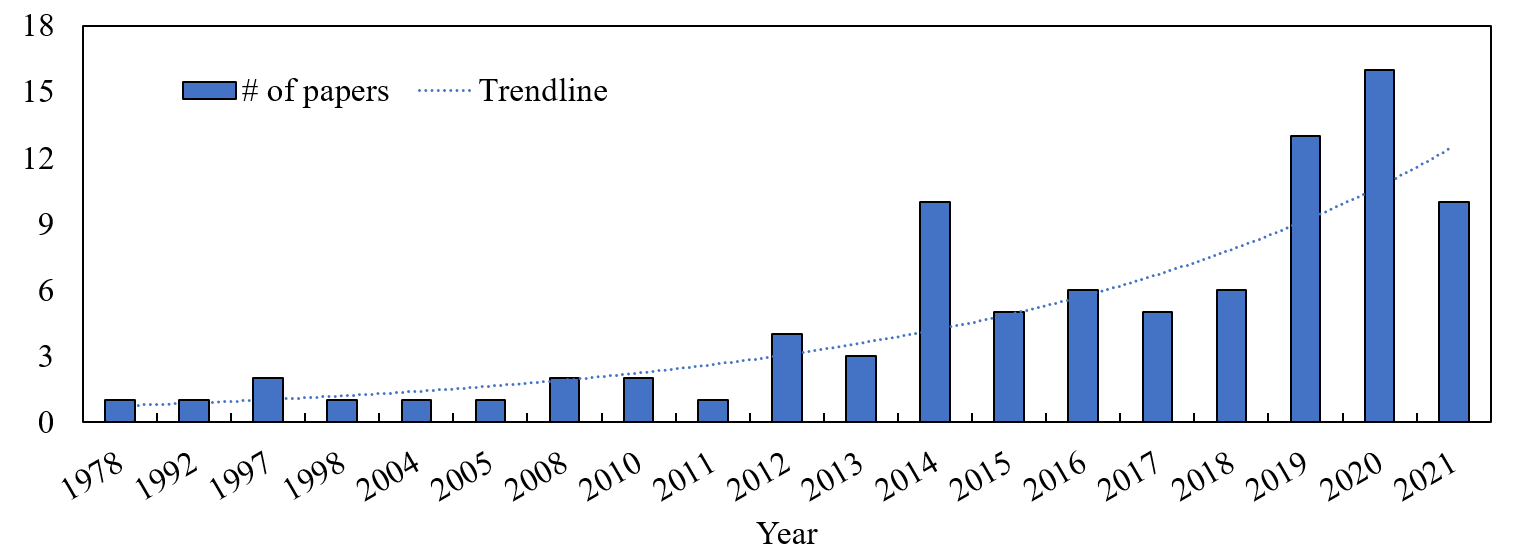}
\caption{Year distribution of reported contributions.}
\label{fig:year}
\end{figure}

To have a broad coverage, we perform a systematic literature survey by indexing papers from several popular digital libraries (IEEE Explore Digital Library, ACM Digital Library, Elsevier ScienceDirect, Springer Digital Library, Google Scholar, Web of Science, DBLP, arXiv) using the keywords "SpGEMM", "sparse matrix", "sparse matrix multiplication", "sparse matrix-matrix multiplication". It is an iterative process, and the keywords are fine-tuned according to the returned results step by step. We try to broaden the search as much as possible while maintaining a manageable result set. Then, we read the titles and abstracts of these papers and finally include 92 SpGEMM related papers, whose year distribution is presented in Figure \ref{fig:year}. It can be seen that the articles in the past three years have shown a rapid upward trend.

It is difficult to give a sufficient and systematic taxonomy for classifying SpGEMM research because the same topic may have different understandings in different contexts, such as load balance of SpGEMM in distributed, shared-memory multicore, single GPU, multi-GPU and CPU+GPU. We select several topics that are frequently covered by existing papers, and present the correlation between papers and topics in table or figure of each section.

The rest of the paper is organized as follows. Section \ref{sec:background} introduces background in detail, including symbol notation used in this paper, some popular and state-of-the-art compression formats for sparse matrix, and several classical applications of SpGEMM. Four different SpGEMM formulations are introduced in Section \ref{sec:formulations}. In Section \ref{sec:keyProblems}, the discussion about the existing solutions for three pivotal problems of SpGEMM computation is presented. In Section \ref{sec:Architecture}, architecture-oriented SpGEMM optimization is introduced. Section \ref{sec:progmmingModel} gives a comprehensive introduction to the SpGEMM optimization using different programming models. Besides, we conduct a performance evaluation covering most SpGEMM implementations in Section \ref{sec:Evaluation}. A discussion about the challenges and future work of SpGEMM is presented in Section \ref{sec:Challenge}, and followed by our conclusion in Section \ref{sec:conclusion}.

\section{Background}\label{sec:background}
\subsection{Preliminaries}\label{sec:Problem}

\subsubsection{Notation}
In this section, we first give the symbolic representation commonly used in this paper, which is presented in Table \ref{tab:notation}. We use bold capital italic for matrices, lowercase bold italic for vectors, and lowercase italic for scalars. 

\begin{table}[h]
\centering
\caption{Symbolic representation.}
\label{tab:notation}
\scalebox{0.75}{
\begin{tabular}{cm{7.3cm}|cm{6.7cm}} 
    \toprule
        \textbf{Symbol} & \textbf{Description} & \textbf{Symbol} & \textbf{Description} \\
    \midrule
        $\matr{A},\matr{B},\matr{C}$ & $\matr{A}$ and $\matr{B}$ are input \deleted{sparse }matrices, $\matr{C}$ is the output \deleted{sparse }matrix & $\cdot$ & Inner product of two vectors\\
    \midrule
        $\vect{a}_{i*}$ & The \deleted{vector of the }$i$-th row of $\matr{A}$ & $\otimes$ & Outer product of two vectors\\
    \midrule
        $\vect{a}_{*j}$ & The \deleted{vector of the }$j$-th column of $\matr{A}$ & $\times$ & General matrix multiplication of two matrices or a matrix and a vector\\
    \midrule
        $\vect{a}_{ij}$ & The \replaced{entry}{element} in the $i$-th row and the $j$-th column of $\matr{A}$ & $*$ & Multiplication of two scalars or a scalar and a vector\\
    \midrule
        $p, q, r$ & \replaced{Dimensions}{The dimension of three matrices}: $\matr{A}: p\times q$, $\matr{B}: q\times r$, $\matr{C}: p\times r$ & $\circ$ & Element-wise multiplication\deleted{ of two matrices} \\
    \bottomrule
\end{tabular}}
\end{table}

\subsubsection{Sparse Matrix}
There is no strict and formal definition of sparse matrix. The most popular one is given by Wilkinson: sparse matrix is any matrix with enough zeros that it pays to take advantage of them \cite{Gilbert1992}. Another common quantitative definition is given by Barbieri et al. \cite{sparse_matrix_definition}, that is, a matrix $\matr{A}$ is sparse if its number of non-zero entries (referred to as NNZ) is $O(n)$.

\begin{table}[h]
\centering
\caption{A summary of sparse formats used in existing contributions.}
\label{tab:sparseFormat}
\scalebox{0.75}{
\begin{tabular}{lm{12.5cm}} 
\toprule
 \textbf{Format} &  \makebox[12cm][c]{\textbf{Contribution}} \\
 \midrule
 COO & \cite{Demirci2019}\cite{Gremse2015}\cite{Matam2012}\cite{Chen2019}\cite{2020_HPCA_SpArch}\cite{2021CiM3D}\cite{2020Mon3D}\cite{Sparta2021}\\
 \midrule
 CSR & \cite{Gustavson1978TwoFA}\cite{Cohen1998}\cite{Buluc2008}\cite{Siegel2010}\cite{Demouth2012}\cite{Matam2012}\cite{Weifeng14}\cite{Patwary2015Parallel}\cite{Liu2015framework}\cite{DaltonOB15}\cite{Kurt2017}\cite{Nagasaka2017}\cite{Elliott2018}\cite{2019_THPC_Performance}\cite{Liu2019}\cite{Winter2019}\cite{2019ACCDNN}\cite{2019GenSpar}\cite{Chen2019}\cite{NAGASAKA2019102545} \cite{2020_HPCA_SpArch}\cite{2020_ICDE_Optimization}\cite{2020_Neural_Sunway}\cite{ishiguro_performance_2020}\cite{2020_PPoPP_spECK}\cite{haghi_fp-amg_2020}\cite{zachariadis_accelerating_2020}\cite{2021optcol}\cite{shivdikar_smash_2021}\cite{zhang_gamma_2021}\cite{hussain2021communication}\cite{xia2021scaling}\cite{rajamanickam2021kokkoskernels}\\
 \midrule
 CSC & \cite{Cohen1998}\cite{zhang_gamma_2021}\cite{zachariadis_accelerating_2020}\cite{rasouli2021compressed}\cite{Chen2019}\cite{2020_ICDE_Optimization}\\
 \midrule
 DCSC & \cite{Buluc2008}\cite{Buluc2008a}\cite{Buluc2010}\cite{BulucG11}\cite{Aydin2011}\cite{Patwary2015Parallel}\cite{AzadBBDGSTW15}\cite{2020MarClu}\\
 \midrule
 Others & DIA\cite{Matam2012},ELL\cite{ishiguro_performance_2020}\cite{Chen2019},DCSR\cite{Buluc2008a},BCSR\cite{Borstnik2014},HNI\cite{2021_AutoRelax},CFM\cite{APPT2019_SPART},Bitmap/BitMask\cite{MICRO2019_SparTen}\cite{MICRO2019_SMASH} \cite{Micro2019_MaxNVM}\cite{zachariadis_accelerating_2020}\cite{HPCA2020_SIGMA},RLC\cite{ISCA2016_EIE}\cite{2017_Eyeriss},RIR\cite{2020CPUFPGA},C$^2$SR\cite{2020_MICRO_MatRaptor},Tiled structure\cite{TileSpGEMM}\\
 \bottomrule
\end{tabular}%
}\end{table}

\subsubsection{Compression Format} 
Storing matrices with a dense pattern often leads to a lot of useless calculations and redundant storage, because they usually have a few non-zero entries\added{ (referred to as non-zeros)}. The prevailing solution is to store each sparse matrix with a compression format. Table \ref{tab:sparseFormat} summarizes the contributions using different \deleted{sparse }formats. 

COO, CSR, CSC, ELL, and DIA are five basic and popular compression formats. COO is the plainest \deleted{compression }format and stores the row index, column index, and value of each non-zero entry in three separate arrays. CSR is the most extensively used \deleted{compression }format in existing work. \deleted{Its only difference from COO is the array storing row information. }Instead of storing row indices, CSR stores the row pointers to the first non-zero entry per row. \deleted{Unlike the row-major format CSR, }CSC \deleted{is a column-major format, which }replaces the column indices array of COO with column pointers. ELL \deleted{format }compacts all \replaced{non-zeros}{non-zero entries} to the left side\deleted{, and stores the column index of each non-zero element}. DIA is \deleted{a compression format }specifically designed for diagonal sparse matrices. It stores \replaced{non-zeros}{non-zero entries} in each diagonal and the offset of each diagonal from the main diagonal. 

In addition to the above five basic formats, some new sparse formats have been proposed over the past years. Bulu{\c{c}} et al. \cite{Buluc2008a}\cite{Buluc2008} propose double compressed sparse column (DCSC), an improved format based on CSC. It is designed for hypersparse matrix by removing all the repetitions in column pointers array. They also present DCSR format \cite{Buluc2008a}, which is a row-based dialect of DCSC. Bor$\check{s}$tnik et al. \cite{Borstnik2014} design an efficient distributed sparse matrix multiplication algorithm using blocked CSR (BCSR) format. Park et al. \cite{2021_AutoRelax} propose huffman-coded non-zero indication (HNI) format, which is a bitmap-based data encoding. It uses non-zero indication bit-stream to replace row and column indices and encodes the stream with Huffman coding. Xie et al. \cite{APPT2019_SPART} design compressed feature map (CFM) for efficiently storing sparse feature maps in convolution neural network (CNN). In SparTen proposed by Gondimalla et al. \cite{MICRO2019_SparTen}, a sparse tensor is encoded into a two tuple of a bit-mask representation and a set of \replaced{non-zeros}{non-zero values}. The bit-mask representation has 1's for positions with \replaced{non-zeros}{non-zero values} and 0's otherwise. This bit-mask based compression idea is also used in \cite{MICRO2019_SMASH}\cite{Micro2019_MaxNVM}\cite{zachariadis_accelerating_2020}\cite{HPCA2020_SIGMA}. Han et al. \cite{ISCA2016_EIE}\cite{2017_Eyeriss} use a CSR variation, run-length encoding (RLE), to store a sparse weight matrix in deep neural network (DNN). The new format stores a vector $\vect{v}$ and an equal-size vector $\vect{z}$ for each column in the weight matrix, where $\vect{v}$ saves the non-zero weights, and $\vect{z}$ stores the number of zeros before the corresponding entry in $\vect{v}$. Soltaniyeh et al. \cite{2020CPUFPGA} propose REAP intermediate Representation (RIP) to increase the throughput on FPGAs. It has three parts: shared feature, metadata and distinct features. To overcome the inefficient memory access of CSR, Srivastava et al. \cite{2020_MICRO_MatRaptor} propose a channel cyclic sparse row (C$^2$SR) format, which assigns each matrix row to a fixed channel in a cyclic manner. In TileSpGEMM, Yu et al. \cite{TileSpGEMM} propose a sparse tile data structure which describes a sparse matrix using two levels of tile information.

\subsection{Typical Applications}\label{sec:Application}
SpGEMM is a basic and critical component in many applications. We introduce the background of some applications and how the SpGEMM is formulated and used in these applications as follows.

\subsubsection{Multi-source BFS}
\begin{figure}[!htbp]
	\centering
	\includegraphics[width=3.8in]{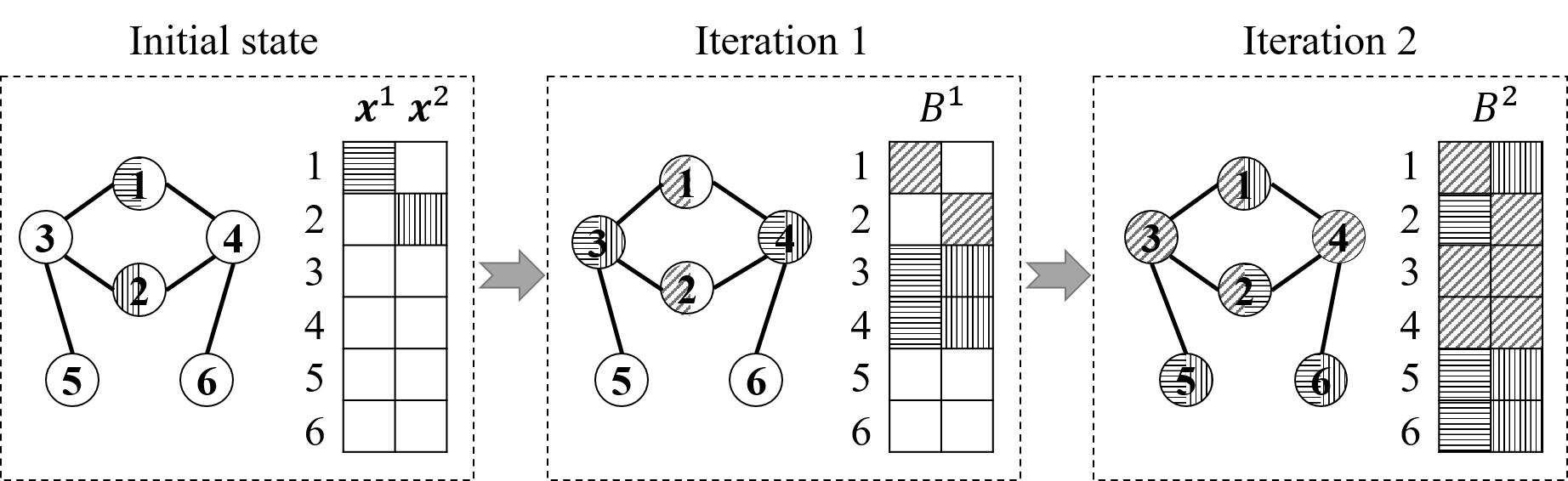}
	\caption{An example of MS-BFS. Vertexes filling in horizontal and vertical lines are being visited by the BFS starting from vertex $\{1\}$ and $\{2\}$ respectively, and vertexes filling in left oblique lines have been visited by the BFS starting from vertex $\{1\}$ or $\{2\}$ \cite{Then2014}.}
	\label{fig:BFS}
\end{figure}

Breadth-first search (BFS) is a key and fundamental subroutine in many graph analysis algorithms\deleted{, such as finding connected domains, finding the shortest path, and finding $k$-hop neighbors \cite{Mattson_GraphPrimitives}\cite{Buluc2010}}. The goal of the BFS is to traverse a graph from a given source vertex, which can be performed by a sparse matrix-vector multiplication (SpMV) between the adjacency matrix $\matr{A}$ of a graph $G=(V,E)$ and a sparse vector representing the source vertex \cite{ZHANG2011}. Assume $n=|V|$, then the size of $\matr{A}$ is $n\times n$. Let $\matr{x}$ be a sparse vector with $\matr{x}_i=1$ and all other entries being zero, then the 1-hop vertexes from source vertex $i$, denoted as $\matr{v}_i^1$, can be derived by the SpMV operation: $\matr{v}_i^1=\matr{A}\times\vect{x}$. Repeating the operation from $\matr{v}_i^1$, we can receive the 2-hop vertexes from $i$. Finally, a complete BFS for the graph $G$ from vertex $i$ is yielded \cite{Gilbert2006}. In contrast, Multi-Source BFS (MS-BFS) runs multiple independent BFSs concurrently on the same graph from multiple source vertexes\replaced{, which can be formulated as SpGEMM.}{. It can be performed by SpGEMM operation.} A simple example for MS-BFS is presented in Figure \ref{fig:BFS}. Let $\{1,2\}$ be two source vertexes, $\matr{A}$ is the adjacency matrix of the graph, and $\matr{X}=(\matr{x}^1,\matr{x}^2)$ is a rectangular matrix representing the source vertexes, where $\matr{x}^1$ and $\matr{x}^2$ are two sparse column vectors with $\matr{x}^1_1=1$ and $\matr{x}^2_2=1$ respectively and all other entries being zero. Then, the sparse matrix $\matr{B}^1$ representing the 1-hop vertexes (denoted as $\vect{v}^1$) from source vertexes $\{1,2\}$ is give by $\matr{B}^1=\matr{A}\times\matr{X}$. Repeat the multiplication of adjacency matrix and $\matr{B}^1$: $\matr{B}^2 = \matr{A}\times \matr{B}^1$, we can derive the 2-hop vertexes from $\{1,2\}$. Finally, we get the results of \replaced{BFS}{breadth-first traversal} from vertices 1 and 2, which are $\{1,3,4,2,5,6\}$ and $\{2,3,4,1,5,6\}$ respectively.

Many applications run hundreds of BFSs over the same graph.\deleted{ Examples of such applications include calculating graph centrality, enumerating the neighborhoods for all vertexes, and solving the all-pairs of the shortest distance problem.} Compared with running BFSs sequentially, running multiple BFSs concurrently in a single kernel allows us to share the computation between different BFSs without paying the synchronization cost \cite{Gilbert2006}. As one of the most expensive operations of MS-BFS, SpGEMM is worthy of more attention to be paid.

\subsubsection{Markov Clustering\added{ (MCL)}}
Clustering is one of the unsupervised learning methods for statistical data analysis, and \replaced{MCL}{Markov clustering (MCL)} is one of the graph clustering algorithms proposed for biological data \cite{2020MarClu}. Using MCL, the closely connected points are grouped into clusters by performing random walks on a graph based on Markov chains. Let $\matr{A}$ denote a probability matrix, and each entry in the matrix is the probability of each point reaching others. The sum of each column in this matrix is 1. There are three steps in the clustering iteration. In the first step of expansion, the probability matrix of reaching other points starting from any point after one step is given by $\matr{B}=\matr{A}\times \matr{A}$. This not only strengthens the connection between different areas, but also leads to the convergence of probabilities. \replaced{In the second step, all entries smaller than a given threshold are pruned.}{Pruning small entries based on a given threshold is applied in the second step to keep the expanded matrix sparse.} Then, the third step of inflation is required to weaken the possibility of loosely connected points by computing the power of each element in the probability matrix. Next, the matrix is replaced with the new matrix and used for next iterations. After a number of iterations, the points in a graph are gradually clustered into groups.\deleted{ The main idea is given in Algorithm \ref{alg:MarkovClustering}.}

Due to the high \added{time and memory }overhead of MCL\deleted{ in time and memory}, high performance MCL (HipMCL) algorithm is proposed for fast clustering of large-scale networks on distributed \replaced{platforms}{storage systems} \cite{2020MarClu}.

\subsubsection{Algebraic Multigrid Solvers}
Algebraic Multigrid (AMG) is a multi-grid method developed based on Geometric multigrid (GMG). AMG iteratively solves large and sparse linear system $\matr{A}\vect{x}=\vect{b}$ by automatically constructing a hierarchy of grids and inter-grid transfer operators \cite{1717312}\cite{Briggs:2000:MTS:347185}. Generally, AMG includes two processes: \textit{setup} and \textit{solve}. The \textit{setup} phase constructs multiple components of multigrid algorithm. The \textit{solve} phase executes multigrid cycling based on these components, and \replaced{SpMV}{sparse matrix-vetor multiplication (SpMV)} dominates this phase. SpGEMM is an important kernel in \textit{setup} phase, whose critical steps are presented in Algorithm \ref{alg:AMG_setup}. It first constructs interpolation operator $\matr{P}_l$ based on input \deleted{sparse }matrix $\matr{A}$ (line 3), then the restriction operator $\matr{R}_l$ is the transposition of $\matr{P}_l$ (line 4). Finally, the coarse-grid system $\matr{A}_{l+1}$ is constructed using Galerkin product (line 5), which is implemented with \replaced{two SpGEMMs}{two sparse matrix-matrix multiplications} \cite{BellDO12}\cite{Ballard2016}. Generally, $\matr{P}_l$ is tall and skinny, and $\matr{R}_l$ is short and fat. \replaced{Their non-zeros' distribution}{The distribution of their non-zeros} is closely related to the used interpolation algorithms.

These \replaced{SpGEMMs}{matrix-matrix multiplications}, taking more than $80\%$ of the total construction time, are the most expensive components. Moreover, the construction of operators (thus SpGEMM) is an expensive part in overall execution since it may occur at every time step (for transient problems) or even multiple times per time step (for non-linear problems), making it important to optimize SpGEMM \cite{Elliott2018}.

\begin{center}
    \scalebox{0.75}{
    \begin{minipage}{0.95\linewidth}
\begin{algorithm}[H]
    \KwIn{$\matr{A}$}
    \KwOut{$\matr{A}_1,...,\matr{A}_L,\matr{P}_0,...,\matr{P}_{L-1}$}
    $\matr{A}_0\leftarrow \matr{A}$\;
    \For{$l=0,...,L-1$}{
        $\matr{P}_l=$interpolation$(\matr{A}_l)$; // Construction of interpolation operator\\
        $\matr{R}_l=\matr{P}_l^T$ // Construction of restriction operator \\
        $\matr{A}_{l+1}=\matr{R}_l\matr{A}_l\matr{P}_l$; // Construction of coarse system: Galerkin product
    }
    \caption{Construction of several important operators in AMG setup phase \cite{BellDO12}.}
    \label{alg:AMG_setup}
\end{algorithm}
\end{minipage}}
\end{center}

\subsubsection{Others}
SpGEMM is also one of the most important components for genome assembly \cite{SC20_Selvitopi}\cite{Guidi2018}\cite{SpGEMM-semiring_2021IPDPS}, NoSQL database \cite{VLDB2009}\cite{Graphulo2015}\cite{Graphulo-HPEC2015}, triangle counting \cite{Cohen2009}\cite{Wolf2015}\cite{Wolf2017}, graph contraction \cite{Gilbert2008}, graph coloring \cite{Kaplan2006}\cite{Deveci2016parallel}, the all pairs shortest path \cite{Chan2007}, sub-graph \cite{Virginia2006}\cite{Buluc2011}, cycle detection or counting \cite{Censor2015}, and molecular dynamics \cite{Weber2015}\cite{Akbudak2014}.\deleted{ Because of limited space, we do not include these applications in detail.}

\section{Formulations}\label{sec:formulations}
\subsection{Overview}

\begin{table}[h]
\centering
\caption{A summary of different SpGEMM formulations.}
\label{tab:formulation}
\scalebox{0.75}{
\begin{tabular}{lm{11cm}}
\toprule
\textbf{Formulation} & \makebox[11cm][c]{\textbf{Contribution}}\\
 \midrule
 Row-by-row & \cite{Gustavson1978TwoFA}\cite{Siegel2010}\cite{Matam2012}\cite{Weifeng14}\cite{Liu2015framework}\cite{Patwary2015Parallel}\cite{DaltonOB15}\cite{Ballard2016}\cite{Nagasaka2017}\cite{DeveciTR17}\cite{Akbudak2018}\cite{Demouth2012}\cite{Deveci2018multithreaded}\cite{Deveci2018multilevelmemory}\cite{Elliott2018}\cite{NAGASAKA2019102545}\cite{2019_THPC_Performance}\cite{Chen2019}\cite{Liu2019} \cite{Winter2019}\cite{2019GenSpar}\cite{2020CPUFPGA}\cite{haghi_fp-amg_2020}\cite{2020_Neural_Sunway}\cite{2020_MICRO_MatRaptor}\cite{2020_TPDS_Cartesian}\cite{shivdikar_smash_2021}\cite{zhang_gamma_2021}\cite{hussain2021communication}\cite{rajamanickam2021kokkoskernels}\cite{Demirci2019}\cite{TileSpGEMM} \\
 \midrule
 Inner-product & \cite{Borstnik2014}\cite{zachariadis_accelerating_2020}\cite{Azad2015}\cite{APPT2019_SPART}\cite{2020MarClu}\cite{Akbudak2018}\cite{Akbudak2017}\cite{2020Mon3D}\\
 \midrule
 Outer-product & \cite{Cohen1998}\cite{YusterR.2004}\cite{Buluc2008}\cite{Buluc2008a}\cite{Matam2012}\cite{Akbudak2014}\cite{2019TACO_MetaStrider}\cite{gu_bandwidth_2020}\cite{Ballard2016}\cite{Akbudak2018}\cite{Akbudak2017}\cite{2020_ICDE_Optimization}\cite{2020_HPCA_SpArch}\cite{Selvitopi2019} \cite{OuterSPACE2018}\\
 \midrule
 Column-by-column & \cite{Buluc2008a}\cite{AzadBBDGSTW15} \\
 \bottomrule
\end{tabular}}
\end{table}

In sparse matrix multiplication, two sparse matrices that are multiplied can be accessed either by row or column. This derives four SpGEMM formulations, which are row-by-row\added{ (RbR)}, \replaced{row-by-column/inner-product (IP)}{inner-product (or row-by-column)}, \replaced{column-by-row/outer-product (OP)}{outer-product (or column-by-row)}, and column-by-column\added{ (CbC)}. Table \ref{tab:formulation} summarizes the existing work using different SpGEMM formulations. As shown in the table, \replaced{RbR}{row-by-row} is the most popular and favored formulation by researchers, followed by \replaced{OP}{outer-product}, then \replaced{IP}{inner-product}, and the least explored is \replaced{CbC}{column-by-column}. In this section, we introduce the calculation of the four formulations, followed by a discussion on the advantages and disadvantages of the four formulations.

\subsection{Row-by-row}
\begin{figure}[!htbp]
	\centering
	\includegraphics[width=1\textwidth]{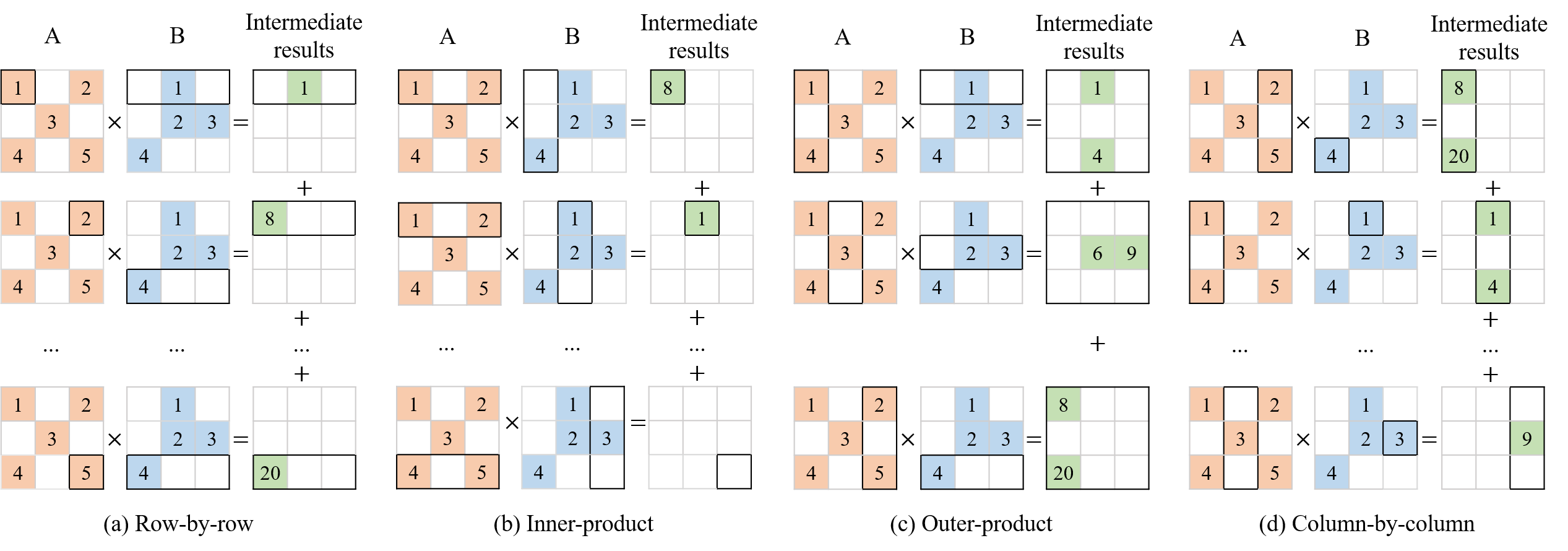}
	\caption{Examples of four SpGEMM formulations. \deleted{The non-zero entries of three sparse matrices are filled with different colors. }The rows, columns or intermediate matrices involved in the computation are labeled with black border \cite{2020_MICRO_MatRaptor}.}
	\label{fig:formulation}
\end{figure}

\replaced{RbR}{Row-by-row} formulation is based on the row-wise partitioning of \added{two }input matrices. Each row $\vect{c}_{i*}$ of $\matr{C}$ is calculated by summing the intermediate multiplication results of each non-zero entry $\vect{a}_{ik}$ of $\vect{a}_{i*}$ and corresponding row $\vect{b}_{k*}$of $\matr{B}$, i.e.
        \begin{equation}
            \vect{c}_{i*}=\sum_{k\in I_i(\matr{A})}\vect{a}_{ik} * \vect{b}_{k*}, i=1,2,...,p.
        \end{equation}
where $I_i(A)$ denotes the set of column indexes $k$ of the \replaced{non-zeros}{non-zero entries} in the $i$-th row of $\matr{A}$. Figure \ref{fig:formulation}(a) presents an example illustrating the computing pattern of \replaced{RbR}{Row-by-row} formulation.

\subsection{Inner-product}
This formulation is based on the row-wise and column-wise partitioning of \deleted{input matrices }$\matr{A}$ and $\matr{B}$ respectively. The result matrix $\matr{C}$ consists of the inner product of each row of \deleted{matrix }$\matr{A}$ and each column of \deleted{matrix }$\matr{B}$, i.e.
        \begin{equation}
            \vect{c}_{ij}=\sum_{k\in I(i,j)}\vect{a}_{ik} * \vect{b}_{kj}, i=1,2,...,p,j=1,2,...,r.
        \end{equation}
where $I(i,j)$ denotes the set of indexes $k$ such that both the entries $\vect{a}_{ik}$ and $\vect{b}_{kj}$ are non-zero. An example that illustrates the inner-product formulation is presented in Figure \ref{fig:formulation}(b).

\subsection{Outer-product}
This formulation is based on the column-wise and row-wise partitioning of input matrices $\matr{A}$ and $\matr{B}$, respectively. The result matrix $\matr{C}$ is calculated by summing the outer product of each column $\vect{a}_{*i}$ of $\matr{A}$ and corresponding row $\vect{b}_{i*}$ of $\matr{B}$, i.e.
        \begin{equation}
            \matr{C} = \sum_{i=1}^{q} \vect{a}_{*i}\otimes \vect{b}_{i*},
        \end{equation}
An example that illustrates the outer-product formulation is presented in Figure \ref{fig:formulation}(c).

\subsection{Column-by-column}
This formulation is based on the column-wise partitioning of two matrices, which is similar to the \replaced{RbR}{row-by-row} formulation. Each column $\vect{c}_{*j}$ of the result matrix $\matr{C}$ is calculated by summing the intermediate multiplication results of each non-zero entry $\vect{b}_{kj}$ of $\vect{b}_{*j}$ and corresponding column $\vect{a}_{*k}$, i.e.
        \begin{equation}
            \vect{c}_{*j}=\sum_{k\in I_j(\matr{B})}\vect{a}_{*k} * \vect{b}_{kj},j=1,2,...,r.
        \end{equation}
where $I_j(\matr{B})$ denotes the set of row indexes $k$ of the \replaced{non-zeros}{non-zero entries} in the $j$-th column of $\matr{B}$. An example that illustrates the column-by-column is presented in Figure \ref{fig:formulation}(d).

\subsection{Discussion}
Srivastava et. al  \cite{2020_MICRO_MatRaptor} compare the data reuse and on-chip memory requirement of four SpGEMM formulations. In their work, data reuse is defined as the ratio of the number of multiply-accumulate (MAC) performed to the size of data read from or written to memory. They assume that all three sparse matrices are square ($N\times N$) and have the uniform non-zero distribution, and $\matr{A}$ and $\matr{B}$ have the same number of non-zero \replaced{entries}{elements} ($nnz$). \deleted{Here, we refer to four formulations as RbR (Row-by-Row), IP (Inner Product), OP (Outer Product), and CbC (Column-by-Column), respectively. }Assuming that the NNZ of $\matr{A}, \matr{B}$, and $\matr{C}$, has a small difference, their discussion can be concluded in two points. First, the relationship between the data reuse of four formulations is: $DR_{IP}<DR_{RbR}=DR_{CbC}<DR_{OP}$. Second, for the on-chip memory, the relationship is $MEM_{IP}<MEM_{RbR}=MEM_{CbC}<MEM_{OP}$.

Here we extend the discussion to the size of intermediate results. Let $nnz_{\matr{C}}$ represent the \replaced{NNZ}{number of non-zero elements} in output matrix $\matr{C}$. For \replaced{RbR}{row-by-row} SpGEMM, one non-zero \replaced{entry}{element} of $\matr{A}$ and the corresponding row of $\matr{B}$ are loaded and multiplied, generating a vector of size $\frac{nnz_{\matr{C}}}{N}$. Therefore, its size of intermediate results that require to be reduced is $O(\frac{nnz\times nnz_{\matr{C}}}{N^2})$. For \replaced{IP SpGEMM}{inner product}, a dot product between one row of $\matr{A}$ and one column of $\matr{B}$ is performed, generating one non-zero \replaced{entry}{element} of $\matr{C}$. Therefore, no intermediate results are generated. For \replaced{OP}{outer-product} SpGEMM, an outer product between one column of $\matr{A}$ and one row of $\matr{B}$ is performed, generating an intermediate result matrix of $\matr{C}$. Its total size of intermediate results is $O(N\times nnz_{\matr{C}})$. \replaced{CbC}{Column-by-column} SpGEMM is similar to \replaced{RbR}{row-by-row} SpGEMM. Therefore, the relationship between the size of intermediate results is $IR_{IP}<IR_{RbR}=IR_{CbC}<OR_{OP}$.

Besides, the four SpGEMM formulations are different in storage format and index matching. RbR prefers to store two input matrices and the output matrix in row-major layout, such as CSR. On the contrary, the column-major layout, such as CSC, is preferred by CbC. However, IP prefers to store two input matrices in row-major and column-major layouts, respectively. OP prefers to an opposite storage format. There is no preference for how the resulting matrix is stored for IP and OP. Among the four SpGEMM formulations, only IP requires index matching.

Generally, the two most basic operations of SpGEMM are scalar multiplication and addition. However, they can also be customized and redefined in some applications. For example, Selvitopi et al. \cite{SC20_Selvitopi} and Guidi et al. \cite{SpGEMM-semiring_2021IPDPS} present a custom semiring to overload \deleted{the numerical }multiplication and addition\deleted{ operators} of SpGEMM in a similar protein sequences identification algorithm. Some popular libraries, such as CombBLAS \cite{CombBLAS}, CTF \cite{CTF-Sparse}, and GraphBLAS \cite{graphblas}, support user-defined multiplication and addition on semirings. On that account, SpGEMM can be generalized and extended to more fields.

\section{Key Problems and Techniques} \label{sec:keyProblems}

\subsection{Overview}
\begin{figure}[!htbp]
    \centering
    \includegraphics[width=1.0\textwidth]{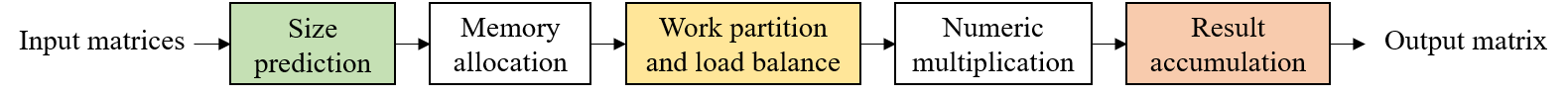}
    \caption{General workflow of SpGEMM.\deleted{ The colored item represents that it is a challenging problem.}}
    \label{fig:overview}
\end{figure}

The typical workflow of SpGEMM is presented in Figure \ref{fig:overview}, which has five stages, including size prediction, memory allocation, work partition and load balance, numeric multiplication, and result accumulation. \replaced{The}{Firstly, the} \textit{size prediction} stage aims to predict the memory footprint of result matrix before real execution. \replaced{The}{Secondly, the} \textit{memory allocation} stage allocates memory space for result matrix on target device. \replaced{The}{Thirdly, the} objective of the \textit{work partition and load balance} stage is to design an efficient algorithm to fully exploit the performance of parallel processors. \replaced{Numeric}{Fourthly, numeric} multiplications and partial additions are performed, and a large number of intermediate results are also generated in this stage. \textit{Result accumulation} desires to reduce these results and calculate final results.

The multiplication result of two sparse matrices is also a sparse matrix, which requires to be stored in a compression format. In CSR/CSC format, NNZ of a sparse matrix dominates its memory footprint, and the sparsity of the result matrix is always unknown in advance. However, precise prediction for the size of result matrix is always expensive in practice. Moreover, with the popularization of multi/many-core processors and distributed systems, the parallelization of intensive computing has become a necessary step for accelerating applications. SpGEMM involves three sparse matrices, which significantly increases the complexity of the problem. Last but not least, due to the sparsity and irregular non-zero distribution, designing an efficient accumulator (the data structure that we use to hold the intermediate results) is also a challenging task.

In the following sections, we discuss in detail the approaches to deal with three challenging problems: size prediction, work partition and load balance, and result accumulation.

\subsection{Size Prediction}\label{sec:SizePrediction}

\subsubsection{\textbf{Precise Prediction}}
\begin{figure}[!htbp]
    \centering
    \includegraphics[width=0.65\textwidth]{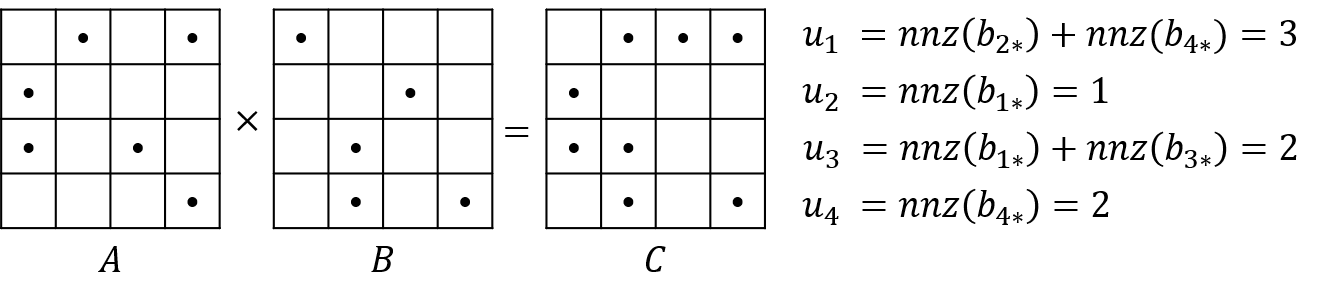}
    \caption{An example of precise and upper-bound prediction. \replaced{Solid dots represent non-zeros}{A solid dot represents a non-zero element}. $u_i$ represents the NNZ in the $i$-th row of $\matr{C}$ predicted by the upper-bound method.}
    \label{fig:symb_phase}
\end{figure}

SpGEMM algorithms using precise prediction usually consist of two phases: symbolic and numeric phases. In the symbolic phase, the precise NNZ in each row/column of the output matrix is computed based on row and column indices of input sparse matrices, an example is shown in Figure \ref{fig:symb_phase}. Real values of \replaced{non-zeros}{non-zero elements} are calculated in the numeric phase.

The implementations of SpGEMM in Kokkos Kernels \cite{rajamanickam2021kokkoskernels}, cuSPARSE \cite{cuSPARSE}, MKL \cite{MKL}, and RMerge \cite{Gremse2015} are typical representatives of this method. Besides, existing work \cite{Demouth2012}, \cite{Nagasaka2017}, \cite{Demirci2019} and \cite{Akbudak2014} also exploit this approach. In order to speed up the symbolic phase, Deveci et al. \cite{DeveciTR17}\cite{Deveci2018multithreaded} design a graph compression technique to compress the matrix $\matr{B}$ by packing its columns as bits. In \cite{gu_bandwidth_2020}, the authors estimate the memory requirement for $\matr{C}$ as well as the number of bins and allocate space for global bins in the symbolic phase. SpECK \cite{2020_PPoPP_spECK} uses size information collected in symbolic execution to guide the selection of accumulators in numeric phase.

\begin{figure}[!htbp]
    \centering
    \includegraphics[width=0.55\textwidth]{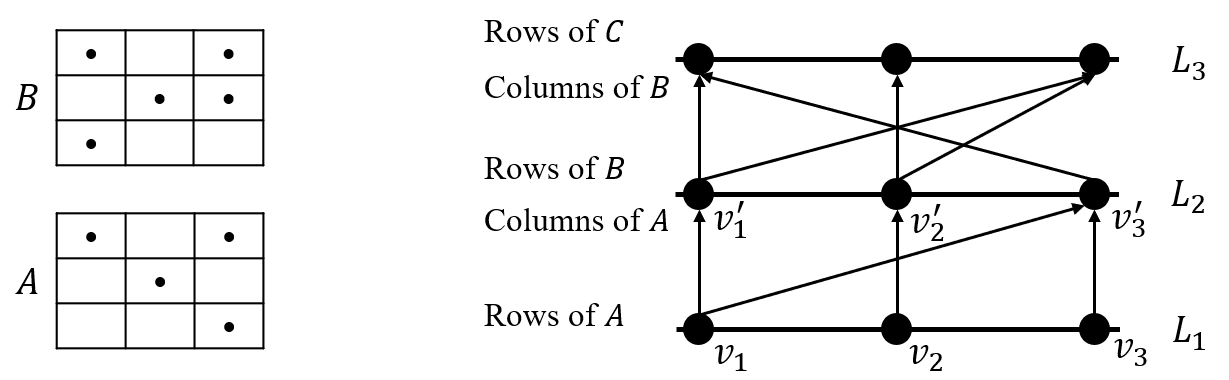}
    \caption{The three-layer graph. Given input matrices $\matr{A}$ and $\matr{B}$ of $3 \times 3$, three nodes ${v_1,v_2,v_3}$ at the layer $L_1$ represent three rows of $\matr{A}$, and  ${{v_1}',{v_2}',{v_3}'}$ at the layer $L_2$ represent its three columns. $v_i$ is connected to ${v_j}'$ only when $a_{ij}$ is non-zero. Then the product $\matr{A} \times \matr{B}$ can be represented as a three-layer graph.}
    \label{fig:three_graph}
\end{figure}

\subsubsection{\textbf{Probabilistic Method}}
Cohen \cite{Cohen1997} transforms the problem of estimating NNZ in the output matrix to the size estimation of reachability sets in a directed graph, and presents a Monte Carlo-based algorithm to estimate the size of reachability sets. The algorithm can be demonstrated using a hierarchical structure graph of matrix product, as shown in Figure \ref{fig:three_graph}. The algorithm starts by assigning a vector of same size, initialized with exponential-distribution random samples, to each node in $L_1$. The vector of each node in the higher layer is equal to the column-wise minimum of the vectors of its neighbors in the lower layer. Finally, NNZ in each row of $\matr{C}$ is estimated based on the vector of each node in $L_3$. For any tolerated relative error $\epsilon>0$, it can compute an $1\pm \epsilon$ approximation of NNZ in result matrix in time $\mathcal{O}(n/\epsilon^2)$. Then, Amossen et al. \cite{AmossenCP14} improve this method to expected time $\mathcal{O}(n)$ for some particular $\epsilon$. Anh et al. \cite{ics_AnhFW16} utilize a similar size estimation technology based on the rows sampling of $\matr{A}$ and $\matr{B}$. In \cite{Cohen1997}, the authors introduce an algorithm to determine the multiplication order of chain products with minimal number of calculation operations. Paper \cite{Selvitopi2019} also uses this three-layer graph representation to estimate the size of result matrix. 

\subsubsection{\textbf{Upper-Bound Prediction}}
The third method computes an upper-bound NNZ in the output matrix and allocates corresponding memory space. The most commonly used method\deleted{ of calculating the upper bound for $\matr{C}$} is to count NNZ in the corresponding rows of $\matr{B}$ for each non-zero entry in $\matr{A}$. Taking the matrices presented in Figure \ref{fig:symb_phase} for example, the upper bound of NNZ in each row of the output matrix is stored in the array $U=\{u_1,u_2,u_3,u_4\}$. The first row of $\matr{A}$ has two \replaced{non-zeros}{non-zero entries}, whose column indices are 2 and 4. Therefore, the upper-bound NNZ in the first row of $\matr{C}$ equals to the sum of NNZ in the second and forth rows of $\matr{B}$. The ESC algorithm \cite{BellDO12}\cite{CUSP} proposed by Bell et al. is a representative of this method. Nagasaka et al. \cite{NAGASAKA2019102545} also count a maximum of scalar non-zero multiplications per row of the result matrix. Then each thread allocates a hash table based on the maximum and reuses the hash table throughout the computation by re-initializing at the beginning of calculating each row.

\subsubsection{\textbf{Progressive Method}}
The fourth method, also known as the progressive method, dynamically allocates memory as needed. It first allocates memory of proper size and then starts matrix multiplication. \replaced{A larger memory block is re-allocated if the current memory is insufficient.}{The reallocation of a larger memory block is required if the current memory is insufficient.} The implementation of SpGEMM in Matlab \cite{Gilbert1992} is a representative of this method. It first guesses the size, and then allocates a memory block that is larger by a constant factor (typically 1.5) than the current space if more space is required at some point. 

Liu and Vinter \cite{Weifeng14}\cite{Liu2015framework} propose a hybrid method which calculates the upper-bound NNZ for each row and groups all rows into multiple bins according to NNZ. The method allocates space of the upper-bound size for short rows and progressively allocates space for long rows. In TileSpGEMM proposed by Yu et al. \cite{TileSpGEMM}, it first calls the symbolic implementation in NSparse \cite{Nagasaka2017} to get the sparse tile structure of $\matr{C}$. Then, it uses binary search to find intersection sparse tiles from $\matr{A}$ and $\matr{B}$. Finally, bit mask operations are used to calculate the number of non-zeros of each tile in $\matr{C}$.

\subsubsection{Discussion}
Of the four methods, precise method not only saves the memory usage, but also enables the sparse structure of $\matr{C}$ to be reused for different multiplies with the same structure of input matrices \cite{Deveci2018multithreaded}\cite{Gustavson1978TwoFA}. Moreover, it presents significant benefits in graph analytics, because most of them work only on the symbolic structure, no numeric phase \cite{Wolf2017}. However, the calculation of two phases means that it needs to iterate through the input matrices twice, leading to higher computation overhead than other methods. The accuracy and the speed of the second method depend on the probabilistic algorithm used, and additional memory allocation must be launched when the estimate fails. The upper-bound method is efficient and easy to implement, but it usually leads to memory over-allocation. The progressive method allocates memory dynamically as needed, and additional memory allocation must also be launched when the first allocation fails. In practice, the choice needs to be made according to the discussed problems.

\subsection{Work Partition and Load Balancing}\label{sec:LoadBalance}
  
    \begin{figure}[h]
    \footnotesize
    \centering
    \begin{displaymath}
    \mbox{Partition}\left\{
    \begin{array}{lr}
    \mbox{Block partition} \left\{
        \begin{array}{lr}
            \mbox{1D} \left\{
                    \begin{array}{lr}
                        \mbox{$T$-partition-\cite{Weifeng14}\cite{Nagasaka2017}\cite{Kurt2017}\cite{DeveciTR17}\cite{Deveci2018multithreaded}\cite{Winter2019}\cite{Elliott2018}\cite{2019GenSpar}\cite{zhang_gamma_2021}\cite{shivdikar_smash_2021}}\\
                        \mbox{$F$-partition-\cite{Azad2015}\cite{5681425}} \\
                        \mbox{$S$-partition-\cite{Deveci2018multilevelmemory}\cite{2020_HPCA_SpArch}\cite{gu_bandwidth_2020}} \\
                        \mbox{Other-\cite{Ballard2016}}\\
                    \end{array}
                \right. \\ 
            \mbox{2D} \left\{
                    \begin{array}{lr}
                \mbox{$TF$-partition-\cite{Buluc2008a}\cite{Buluc2008}\cite{Jin2004}\cite{Patwary2015Parallel}} \\
                        \mbox{$TS$-partition-\cite{Deveci2018multilevelmemory}} \\
                    \end{array}
                \right. \\
             \mbox{3D-\cite{AzadBBDGSTW15}\cite{hussain2021communication}\cite{Weber2015}\cite{TileSpGEMM}} \\
        \end{array}
    \right. \\ 
    \mbox{Graph partition} \left\{
        \begin{array}{lr}
            \mbox{Hypergraph partition-\cite{Akbudak2014}\cite{Ballard2016hypergraph}\cite{Kurt2017}\cite{Akbudak2017}\cite{Akbudak2018}\cite{Selvitopi2019}\cite{2020_TPDS_Cartesian}} \\
            \mbox{Bipartite graph partition-\cite{Akbudak2018}\cite{Demirci2019}\cite{Selvitopi2019}}\\
        \end{array}
    \right. \\ 
    \end{array}
    \right.
    \end{displaymath}
    \caption{Classification of SpGEMM partition.}
    \label{fig:matrix_partition}
\end{figure}

The mainstream work partition and load balancing methods are presented in Figure \ref{fig:matrix_partition}, and we introduce and discuss these methods in detail in the following sections.

\subsubsection{Block Partition}
\begin{figure}[htbp]
    \centering
    \includegraphics[width=0.75\textwidth]{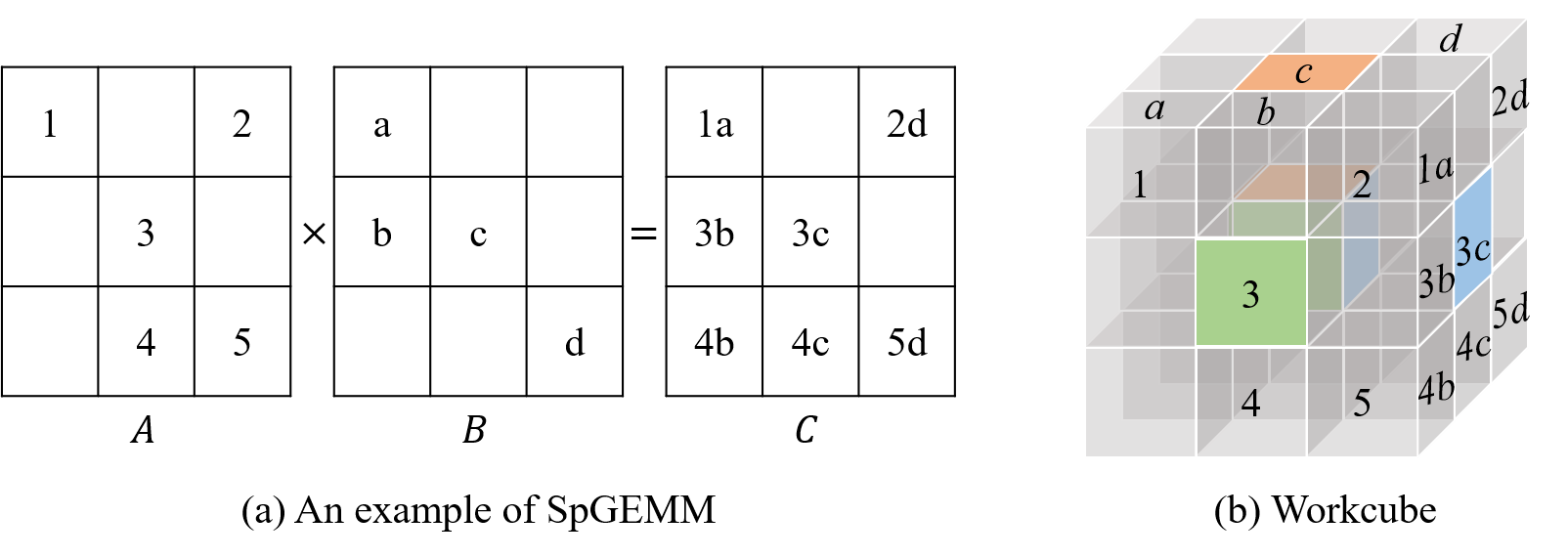}
    \caption{The workcube for matrix multiplication.}
    \label{fig:workcube}
\end{figure}

Block partition of SpGEMM can be categorized into 1D, 2D and 3D algorithms based on how they partition the work among computing units \cite{AzadBBDGSTW15}\cite{Ballard2016hypergraph}. We first introduce the workcube notation before diving into the details, as shown in Figure \ref{fig:workcube}(b). The front, top, and side views of the workcube represent the two input matrices and the output matrix presented in Figure \ref{fig:workcube}(a), respectively. The workcube can be divided into $3\times 3\times 3$ voxels. Each voxel inside represents the scalar multiplication of two non-zeros, which are mapped to cells of the voxel in the front and top views, and contributes to its projection in the side view \cite{Ballard2013_communication}. Based on this projection, task partition of SpGEMM can be seen as the partition of workcube in different dimensions. Next, we introduce 1D, 2D and 3D partition based on this notation.

\begin{figure}
    \centering
    \includegraphics[width=0.7\textwidth]{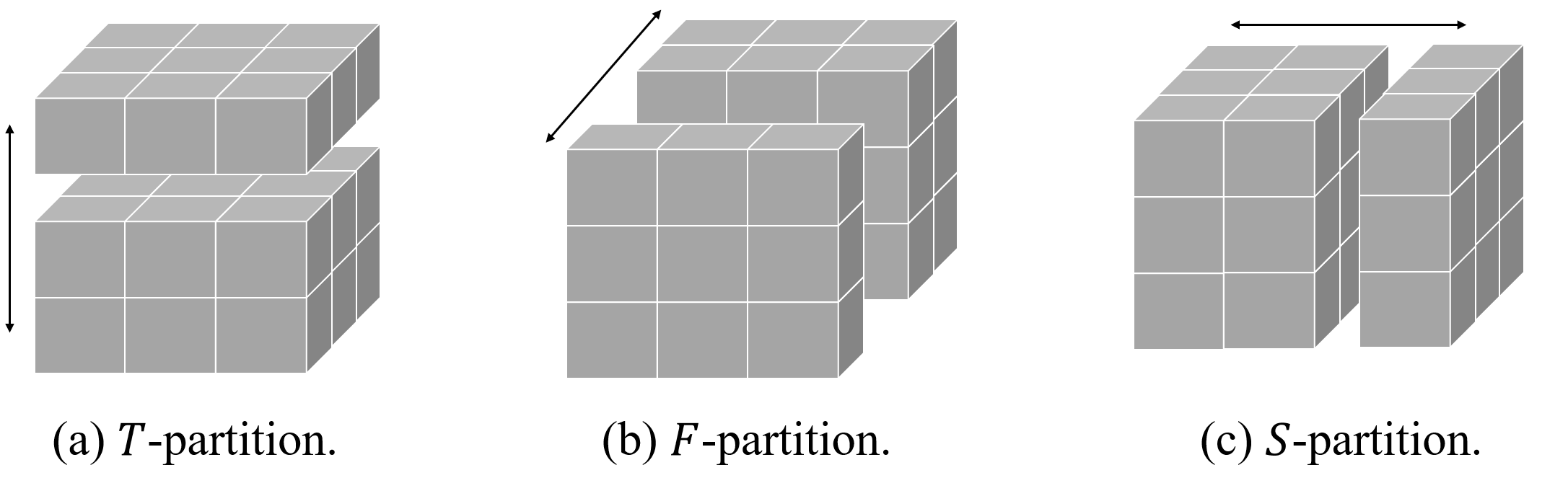}
    \caption{Three variants of 1D partition.}
    \label{fig:1D}
\end{figure}

\textbf{1D Partition}. 1D partition only divides the workcube in one of the three dimensions. As shown in Figure \ref{fig:1D}, each sub-figure represents a variant of 1D partition. A "layer" of the workcube is assigned to one computing unit in all variations. Figure \ref{fig:1D}(a) divides the workcube by planes parallel to the top view, referred to as $T$-partition. Each computing unit is responsible for the multiplication of a set of rows in $\matr{A}$ and the entire matrix $\matr{B}$, producing the corresponding rows of $\matr{C}$. Similarly, Figures \ref{fig:1D}(b) and (c) divide the workcube by planes parallel to the front view and side view, respectively. We refer to these two partition variants as $F$-partition and $S$-partition.

\textbf{$T$-partition}. Liu et al. \cite{Weifeng14} group rows of $\matr{C}$ to different bins according to their upper-bound NNZ to maintain a balanced load. Nagasaka et al. \cite{Nagasaka2017} partition rows of $\matr{C}$ according to precise NNZ in each row in numeric phase. Kurt et al. \cite{Kurt2017} propose to assign one thread block to a fixed number of rows in $\matr{A}$ and assign a same number of threads in the block to complete the row-wise multiplication of each row in $\matr{A}$ and corresponding entries in $\matr{B}$. Deveci et al. \cite{DeveciTR17} propose a hierarchical and parallel SpGEMM algorithm based on Kokkos library \cite{Kokkos}. At the first level, one \textit{kokkos}-\textit{team} is assigned to calculate a set of rows of $\matr{C}$. Each \textit{kokkos}-\textit{thread} within the team is responsible to produce a subset of these rows at the second level. Multiple vector-lanes in one \textit{kokkos}-\textit{thread} are assigned to perform the row-wise multiplication of each non-zero entry in the subset of $\matr{A}$ and corresponding rows in $\matr{B}$ at the third level. They also explore $T$-partition but with different task assigning schemes for \replaced{HPC}{high performance computing} architectures \cite{Deveci2018multithreaded}. Winter et al. \cite{Winter2019} present a partition scheme assigning the same NNZ of $\matr{A}$ to each block while ignoring the row boundaries\deleted{ of a matrix}. Instead of splitting rows evenly \cite{Elliott2018}, Li et al. \cite{2019GenSpar} split the matrix into multiple row blocks based on NNZ. To address the input or output reuse problem, Zhang et al. \cite{zhang_gamma_2021} split matrix $\matr{A}$ into row fibers and dispatch them to processing elements (PEs) in a SpGEMM accelerator GAMMA, and each PE then performs a linear combination of row fibers of $\matr{B}$ to produce a row fiber of \deleted{output matrix }$\matr{C}$. Shivdikar et al. \cite{shivdikar_smash_2021} group multiple rows in a single window and assign one PIUMA (Programmable Integrated Unified Memory Architecture) block to a window. The size of a window depends on the scratchpad size.

\textbf{$F$-partition}. Azad et al. \cite{Azad2015} assign processors to process columns of the upper triangular matrix in triangle counting. Lin et al. \cite{5681425} design an architecture for SpGEMM on FPGAs, the computation is partitioned evenly to all PEs, and each PE is assigned to calculate multiple columns of matrix $\matr{C}$. 

\textbf{$S$-partition}. Based on the outer-product multiplication, Deveci et al. \cite{Deveci2018multilevelmemory} divide rows of $\matr{B}$ into blocks so that each block can be fitted into the HBM of KNL. This partition also induces a row-wise partition of $\matr{A}$. Zhang et al. \cite{2020_HPCA_SpArch} optimize the outer-product SpGEMM by compacting non-zeros of each row in $\matr{A}$ to the left so that the data locality for both input and output matrices are jointly optimized. Gu et al. \cite{gu_bandwidth_2020} develop an improved ESC SpGEMM based on outer product. They group the expanded triples into bins to saturate memory bandwidth. 

\textbf{Others}. Ballard et al. \cite{Ballard2016} present a theoretical and detailed analysis for the communication cost of Galerkin triple product in the smoothed aggregation of AMG methods. They conclude that the row-by-row product ($T$-partition) is the best 1D method for the first two multiplications, and the outer product ($S$-partition) is the best 1D method for the third multiplication. 

Most of SpGEMM can be implemented using 1D partition. However, when the NNZ in columns or rows of the matrix $\matr{A}$ or $\matr{B}$ increases, the communication overhead becomes substantially huge in distributed computing. This indicates that a "layer" in 1D partition is required to be processed by more computing units. Therefore, the 2D algorithms with fine-grained partition are proposed.

\begin{figure}[!htbp]
    \centering
    \includegraphics[width=0.75\textwidth]{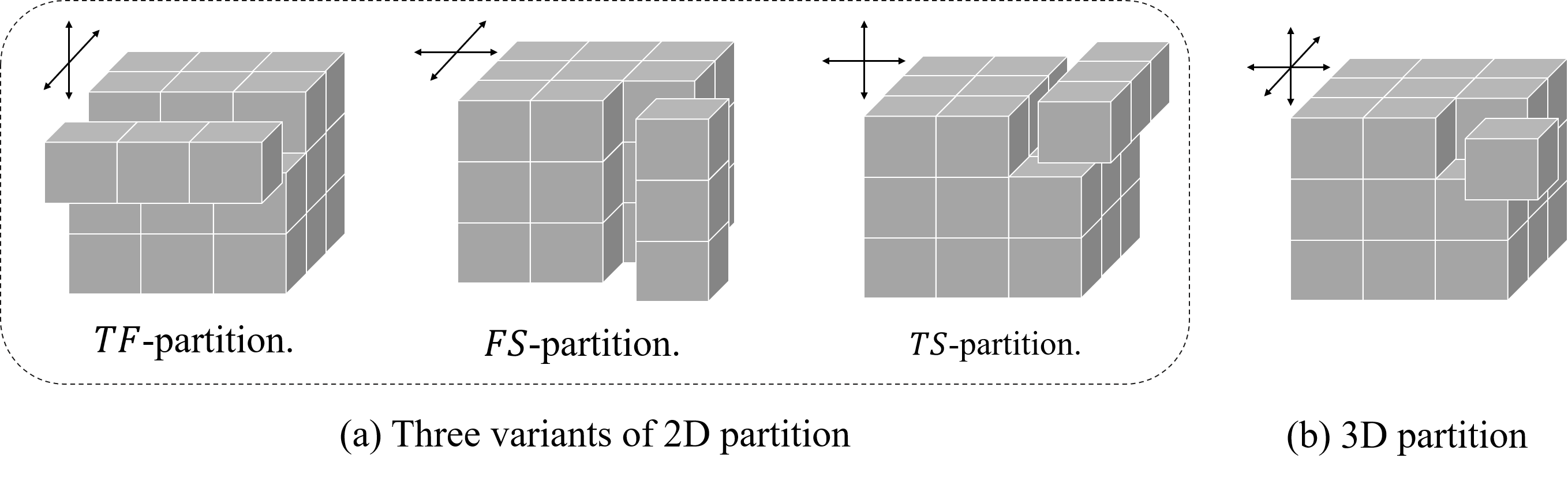}
    \caption{\replaced{2D and 3D partition.}{Three variants of 2D partition.}}
    \label{fig:2D}
\end{figure}

\textbf{2D Partition}. 2D algorithms divide the workcube in two of the three dimensions, so there are also three variants\replaced{. Figure \ref{fig:2D}(a) illustrates}{, as shown in Figure \ref{fig:2D}. Figure \ref{fig:2D}(a), (b) and (c) illustrate} three partition variants that divide the workcube by planes parallel to two of the three views (top and front views, front and side views, top and side views), referred to as $TF$-partition, $FS$-partition and $TS$-partition hereafter. In $TF$-partition, each computing unit computes a block of \deleted{the output matrix }$\matr{C}$. However, in $TS$-partition and $TF$-partition, each computing unit computes a block of intermediate results of $\matr{C}$. 

\textbf{$TF$-partition}. 
SpSUMMA was first proposed by
Bulu{\c{c}} et al. \cite{Buluc2008a}\cite{Buluc2008} based on dense SUMMA algorithm \cite{vandeGeijn1995}. They also present a comparative SpCannon algorithm based on dense Cannon algorithm \cite{Cannon1969}. Both algorithms logically organize processors as a 2D grid, and map a block of $\matr{A}$ and $\matr{B}$ to each processor. In SpSUMMA, each processor broadcasts data to other processors in the same row/column of the grid, and also receives data broadcast by these processors. In SpCannon, processors exchange data using point-to-point communication. In both algorithms, each processor accesses a row block of $\matr{A}$ and column block of $\matr{B}$, and calculates a result block of $\matr{C}$, which conforms with $TF$-partition. Matrix partition similar to SUMMA is also used by Jin et al. \cite{Jin2004}. Patwary et al. \cite{Patwary2015Parallel} partition \deleted{matrix }$\matr{A}$ by row, while partition \deleted{matrix }$\matr{B}$ by column when a certain condition is satisfied.

\textbf{$TS$-partition}. Deveci et al. \cite{Deveci2018multilevelmemory} partition $\matr{A}$ and $\matr{B}$ by row for GPU. When one partition of $\matr{A}$ cannot fit in GPU fast memory, column-wise partition is applied for row strip of $\matr{A}$.

2D algorithms perform reasonably well on a few hundred processes. However, as the number of processes increases, the communication cost becomes a bottleneck, and 3D algorithms were developed to reduce the communication cost.


\textbf{3D Partition}.
3D algorithms divide all three dimensions of the workcube. As shown in Figure \replaced{\ref{fig:2D}(b)}{\ref{fig:3D}}, each computing unit owns a block of $\matr{A}$ and a block of $\matr{B}$ to compute an intermediate result block of the matrix $\matr{C}$.

Azad et al. \cite{AzadBBDGSTW15} present a parallel implementation of the 3D SpGEMM algorithm. Hussain et al. \cite{hussain2021communication} use a similar 3D partition, while dividing each block of $\matr{B}$ into multiple batches to meet the size of available memory. To avoid excessive communications and logistic operations of 2D SpSUMMA, Weber et al. \cite{Weber2015} propose a novel SpGEMM algorithm MPSM3 for the computation of the density matrix in electronic structure theory. All processes are organized as a 3D Cartesian topology, and each box in the workcube is assigned to a specific process on terms of the physical properties of the problem. Yu et al. \cite{TileSpGEMM} divide two input sparse matrices into a number of 16-by-16 tiles to utilize 8-bit unsigned char data type for local indices.

\subsubsection{Graph Partition} 
Traditional block-based matrix partitions rarely consider the workload (NNZ in each block) assigned to each processing element. For example, SpSUMMA presumes that non-zeros are uniformly and randomly distributed across the row/column, which may not hold for most sparse matrices. 

\begin{figure}[htp]
    \centering
    \includegraphics[width=0.75\textwidth]{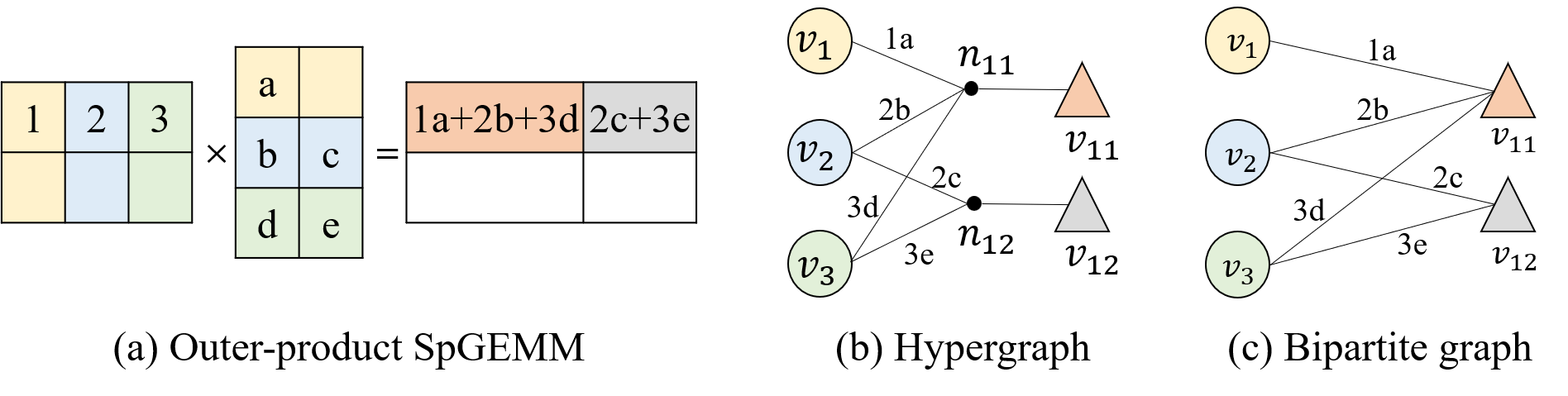}
    \caption{Hypergraph and bipartite graph representation of SpGEMM.}
    \label{fig:hypergraph}
\end{figure}

\textbf{Hypergraph Partition}. The hypergraph representation for $\matr{C}=\matr{A}\times \matr{B}$, denoted as $\mathcal{H}=(\mathcal{V},\mathcal{N})=(\mathcal{V}^{\matr{AB}}\cup \mathcal{V}^{\matr{C}},\mathcal{N})$, is defined as a set of vertices $\mathcal{V}$ and a set of nets (hyperedges) $\mathcal{N}$. Here we take the outer-product SpGEMM as an example to introduce the hypergraph based partition. In $\mathcal{H}=(\mathcal{V}^{\matr{A}\matr{B}}\cup \mathcal{V}^{\matr{C}},\mathcal{N})$, each vertex $v_i$ in $\mathcal{V}^{\matr{AB}}$ denotes the outer product of the column $\vect{a}_{*i}$ of $\matr{A}$ with the row $\vect{b}_{i*}$ of $\matr{B}$. $\mathcal{V}^{\matr{C}}$ contains a vertex $v_{ij}$ for each non-zero entry $\vect{c}_{ij}$ in $\matr{C}$. $\mathcal{N}$ contains a net $n_{ij}$ for each non-zero entry $\vect{c}_{ij}$ in $\matr{C}$. Figure \ref{fig:hypergraph}(a) shows an example of SpGEMM and Figure \ref{fig:hypergraph} (b) shows its hypergraph representation. The task of hypergraph partitioning is to divide a hypergraph into two or more roughly equal-sized parts such that a cost function on the nets connecting vertices in different parts is minimized \cite{2011HP-book}.

\textbf{Bipartite Graph Partition}. 
A bipartite graph for $\matr{C}=\matr{A}\times \matr{B}$, denoted as $\mathcal{G}=(\mathcal{V}^{\matr{A}\matr{B}}\cup \mathcal{N}^{\matr{C}}, \mathcal{E})$, is defined as two disjoint sets of vertices $\mathcal{V}^{\matr{A}\matr{B}}$ and $\mathcal{V}^{\matr{C}}$, and a set of edges $\mathcal{E}$. Also take outer-product SpGEMM as an example. The semantics of each vertex in $\mathcal{G}$ is the same with those in $\mathcal{H}$. The difference is that the dependence of $\vect{c}_{ij}$ is captured with edges, instead of a net. If an outer product represented by $v_x$ produces a partial result for $\vect{c}_{ij}$, an edge connecting vertices $v_x$ and $v_{ij}$ is added in $\mathcal{E}$. An example of a bipartite graph for the outer-product SpGEMM is presented in Figure \ref{fig:hypergraph}(c).

Hypergraph-based partition was first applied in outer-product SpGEMM by Akbudak et al. \cite{Akbudak2014}. Besides, they also present two extended hypergraph model, which partitions the matrix $\matr{C}$ by row and column, respectively. Ballard et al. \cite{Ballard2016hypergraph} present a fine-grained hypergraph model, in which each vertex is either a scalar multiplication or a non-zero entry. Based on this hypergraph model, Kurt et al. \cite{Kurt2017} use an iterative method to construct hypergraph partitions to ensure that the data entries involved in each partition does not exceed the cache capacity. Akbudak et al. \cite{Akbudak2017} use hypergraph model to improve outer-product and inner-product SpGEMM.  

Akbudak et al. \cite{Akbudak2018} propose and compare multiple
computational partitioning models based on hypergraph and bipartite graph partitioning, with the aim of reducing message volume. Further, three communication hypergraph partitioning models for three SpGEMM formulations are proposed to reduce the latency cost. The bipartite graph model for row-by-row SpGEMM was later used by Demirci et al. \cite{Demirci2019}. Instead of associating two weights with each vertex, they propose a three-constrained partitioning in which each vertex is associated with three weights. Selvitopi et al. \cite{Selvitopi2019} apply bipartite graph and hypergraph models for simultaneous scheduling of the map and reduce tasks for MapReduce jobs. Demirci et al. \cite{2020_TPDS_Cartesian} propose hypergraph models for 2D and 3D partitioning to improve the performance of 2D and 3D SpGEMM. 

\subsubsection{Discussion}
Each partition algorithm usually has its own design considerations and advantages. As the most commonly used \replaced{sparse formats}{compression formats for sparse matrices}, CSR and CSC usually store sparse matrices by rows and columns, respectively. It is compatible with the 1D partition. The workload of 1D partition is closely related to the non-zeros distributions in the divided rows or columns, and the non-zeros distributions in the other sparse matrix. Taking $T$-partition for example, we assume that each thread or a group of threads are responsible for the multiplication of one row of $\matr{A}$ and entire $\matr{B}$. If $\matr{B}$ is a regular sparse matrix and has uniform non-zeros distribution, then the NNZ in each row of $\matr{A}$ determines the workload of each computing unit. Therefore, the longest row of $\matr{A}$ dominates the performance. The problem of load imbalance is more serious and complicated if $\matr{B}$ is irregular. 2D and 3D partitions are fine-grained and produce a more balanced workload than the 1D partition. Communication cost is an important metric for the distributed system. In the 1D partition, each work process accesses an entire input matrix ($\matr{B}$ for $T$-partition, $\matr{A}$ for $F$-partition) or output matrix ($\matr{C}$ for $S$-partition), which requires a huge communication overhead. Differently, 2D and 3D partitions just access some rows/columns, or partial \replaced{non-zeros}{non-zero elements}. The graph partitioning algorithms build computation and communication graph models to ensure load balance and minimum communication cost. For SpGEMM with an irregular computation pattern, graph partitioning algorithms achieve a more balanced load and lower communication cost than block partitioning algorithms, but at the cost of higher partitioning overhead.

\subsection{Result Accumulating}\label{sec:ResultMergin}
According to the storage structure of the intermediate results, \replaced{we classify accumulators}{accumulators can be classified} into three types: dense, sparse, and hybrid accumulators. The \replaced{related researches are}{research on each accumulator is} shown in Figure \ref{fig:accumulators}. 

\begin{figure}[!ht]
    \footnotesize
    \centering
    \begin{displaymath}
    \mbox{Accumulators}\left\{
    \begin{array}{lr}
    \mbox{Dense-\cite{Gilbert1992}\cite{Gustavson1978TwoFA}\cite{Siegel2010}\cite{Patwary2015Parallel}\cite{Elliott2018}\cite{2020_ICDE_Optimization}} \\ 
    \mbox{Sparse} \left\{
        \begin{array}{lr}
            \mbox{List-based} \left\{
                    \begin{array}{lr}
                        \mbox{Radix sort-\cite{BellDO12}\cite{DaltonOB15}\cite{gu_bandwidth_2020}\cite{Liu2019}} \\
                        \mbox{Merge sort-\cite{Gremse2015}\cite{Liu2019}} \\
                        \mbox{Heap sort-\cite{AzadBBDGSTW15}\cite{NAGASAKA2019102545}} \\
                    \end{array}
                \right. \\ 
            \mbox{Hash-based-\cite{Deveci2013}\cite{ics_AnhFW16}\cite{Nagasaka2017}\cite{shivdikar_smash_2021}\cite{Yasar2018}\cite{Liu2019}\cite{Weber2015}\cite{Winter2019}\cite{2020MarClu}}\\
        \end{array}
    \right. \\ 
    \mbox{Hybrid-\cite{DeveciTR17}\cite{Deveci2018multithreaded}\cite{2020_PPoPP_spECK}\cite{2019GenSpar}\cite{Yasar2018}\cite{Weifeng14}\cite{Liu2015framework}\cite{TileSpGEMM}}\\ 
    \end{array}
    \right.
    \end{displaymath}
    \caption{Classification of accumulators.}
    \label{fig:accumulators}
\end{figure}

\subsubsection{Dense Accumulator}
\begin{figure}[htp]
    \centering
    \includegraphics[width=0.75 \textwidth]
    {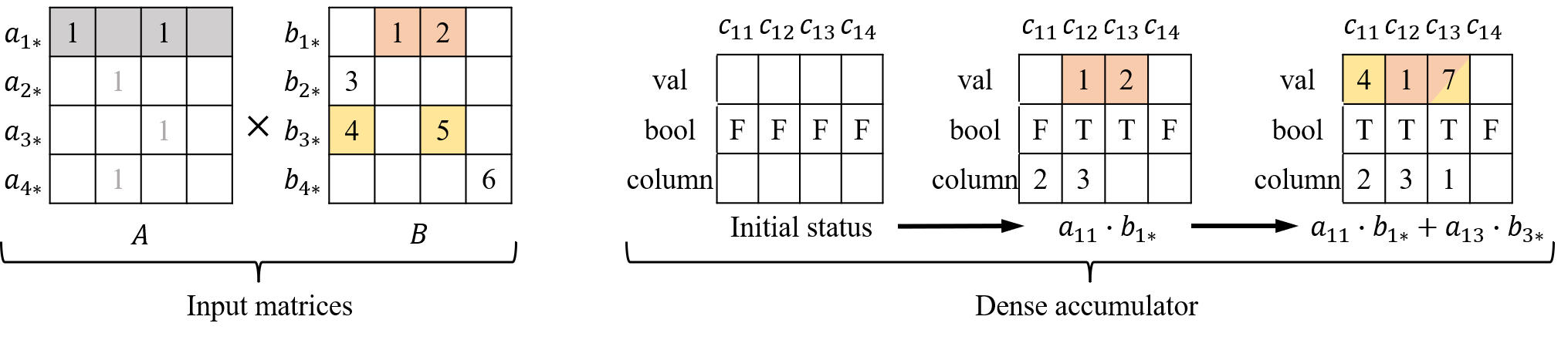}
    \caption{An example for dense accumulator. Each non-zero entry in the first row of $\matr{A}$ is multiplied by the corresponding row of $\matr{B}$. The intermediate results are stored in the dense array $val$, the corresponding entry in the array $bool$ is set to true, and the column index that has not been recorded is pushed into the array $col$.}
    \label{fig:dense_accumulators}
\end{figure}

Dense accumulators use dense vectors to cache the intermediate results of current "active" columns or rows in the result matrix. The most popular dense accumulator, was first proposed by Gustavson \cite{Gustavson1978TwoFA}, usually has three vectors. The first one stores real values. The second one is used as a tag array to check if a column index has been inserted before. The third one stores column indexes\deleted{ of non-zero entries}. An example of the dense accumulator is presented in Figure \ref{fig:dense_accumulators}.

Gustavson's algorithm allows random access to a single \replaced{entry}{element} in a specified time, and it is widely used and improved by many successive researches \cite{Gilbert1992}\cite{Patwary2015Parallel}\cite{Siegel2010}\cite{2020_ICDE_Optimization}. For example, the SPA accumulator in Matlab \cite{Gilbert1992} uses a dense vector of the same size as the column size of $\matr{C}$, a dense vector with true/false "occupied" flags, and an unordered list of the indices whose "occupied" flags are true. Elliott et al. \cite{Elliott2018} note that prolongation matrices in the AMG method are usually tall and skinny, so they use a local dense lookup array for each thread that supports fast memory space allocation for result matrix by page-aligned and size-tracked techniques.

\subsubsection{Sparse Accumulator} In sparse accumulator, intermediate results for a row are stored in compact data structures. There are two types of method for \deleted{intermediate result }merging: list\deleted{-based} and hash-based methods.

\textbf{List-based} accumulating usually sorts the intermediate results \replaced{by column indexes}{according to column indexes using different sorting algorithms}. According to the sorting algorithms utilized, we classify researches as follows. \textbf{Radix sort} based methods allocate the entries to be sorted to some "buckets", so as to achieve the goal of sorting. Bell et al. \cite{BellDO12} propose the ESC algorithm, which sorts many scalar products in the second phase. It has been improved in \cite{Liu2015framework} and \cite{2019GenSpar}. Dalton et al. \cite{DaltonOB15} use the B40C radix sort algorithm \cite{Merrill2011} which allows specifications in the number and location of the sorting bits to accelerate the sort operation. Gu et al. \cite{gu_bandwidth_2020} use an in-place radix sort to group keys by a single byte sharing the same valid byte position. In \cite{Liu2019}, a sparse accumulator, which sorts data in the GPU registers, is proposed. \textbf{Merge sort} based methods first divide the sequence to be sorted into several subsequences. Then each subsequence is ordered, and all subsequences are combined into an overall ordered sequence. In \cite{Gremse2015}, the sparse rows of $\matr{B}$ selected and weighted by corresponding non-zeros of $\matr{A}$, are merged in a hierarchical way similar to merge sort. Besides, Liu et al. \cite{Liu2019} propose a GPU register-based merge algorithm, which demonstrates significant speedup over its original implementation. \textbf{Heap sort} based accumulator is proposed by Azad et al. \cite{AzadBBDGSTW15}. The intermediate result is represented as a list of triples, and each triple includes row index, column index and value of each non-zero entry. A $k$-way merge on $k$ lists of triples is performed by maintaining a heap of size $k$, which stores the current minimum entry in each triple list. Then a multi-way merge routine finds the minimum triple from the heap and merges it into the current result. Nagasaka et al. \cite{NAGASAKA2019102545} implement a heap-sort based sparse accumulator on multicore architectures.

\begin{figure}[!htbp]
    \centering
    \includegraphics[width=0.7\textwidth]
    {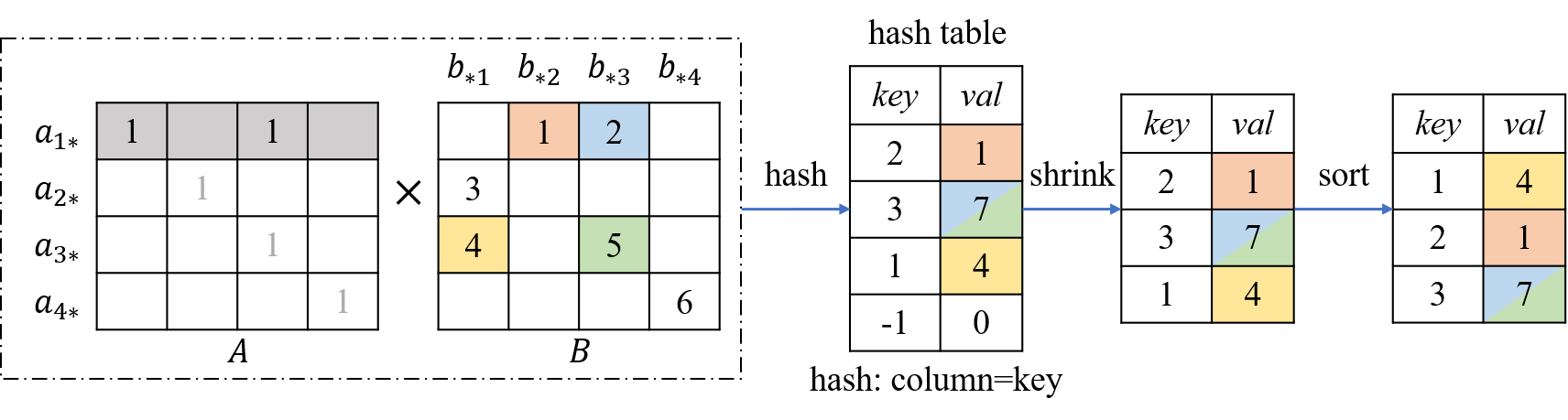}
    \caption{An example of hash-based accumulator for calculating the first row of the output matrix $\matr{C}$. The hash table is initially set to -1. The column indices of $\matr{B}$ serve as the key of hash table.}
    \label{fig:hash_accumulators}
\end{figure}

\textbf{Hash-based.} Hash-based sparse accumulator first allocates a memory space based on the upper bound estimation as the hash table and uses the column indexes of the intermediate results as the key. Then the hash table is required to be shrunk to a dense state. Finally, we sort the values of each row of the result matrix according to their column indexes to obtain the final result matrix compressed with sparse format \cite{Liu2019}. An example is shown in Figure \ref{fig:hash_accumulators}. SpGEMM implementation in cuSPARSE \cite{Demouth2012}\cite{cuSPARSE} is a representative of this accumulator. Deveci et al. \cite{Deveci2013} design a hashmap-based, two-level, sparse accumulator that supports parallel insertions and merges from multiple vector lanes. Weber et al. \cite{Weber2015} use the hash table to store each row of the matrix $\matr{C}$ based on the distributed BCSR format. In \cite{ics_AnhFW16}, the row and column indices of an intermediate result, normally stored in two 32-bit numbers, are first compacted into a single 32-bit value and then inserted into hash table as key. It is helpful to accelerate the later sorting operation. Nagasaka et al. \cite{Nagasaka2017} utilize vector registers in KNL and multi-core architectures for hash probing to accelerate hash-based SpGEMM. Their hash function is defined as the reminder after division of the column index multiplied by a constant number by hash table size. Liu et al. \cite{Liu2019} optimize the data allocations of each thread in a hash-based accumulator utilizing GPU registers. Selvitopi et al. \cite{2020MarClu} compare the heap \deleted{sort }and hash sort in distributed HipMCL, and conclude that the latter method performs better.

\subsubsection{Hybrid Accumulator}
Deveci et al. develop KKMEM \cite{DeveciTR17} and KKSpGEMM \cite{Deveci2018multithreaded} algorithms, which support two hash-based accumulators and one dense accumulator. Which accumulator to use depends on the features of the discussed matrix. KKTri-Cilk algorithm \cite{Yasar2018} uses a similar accumulator to \added{that in }the KKSpGEMM algorithm. Meanwhile, it is optimized in the Cilk implementation.

Sparse or dense accumulators are selected according to the distribution of \replaced{non-zeros}{non-zero entries} in different parts of a matrix. Liu et al. \cite{Weifeng14}\cite{Liu2015framework} propose a hybrid parallel result accumulating method. Three sort-based accumulators are selected according to the upper-bound NNZ in each row. In the spECK algorithm \cite{2020_PPoPP_spECK}, a local balancer decides which accumulator to use for each block to achieve the best expected performance from the following three accumulators: direct reference, hashing, or dense accumulation. The hash function of its hash-based accumulator multiplies the element index with a prime number and then divides the result by hash table size. TileSpGEMM \cite{TileSpGEMM} uses the sparse accumulator working on a sparse tile and the dense accumulator working on a dense tile. Whether a tile is dense or sparse is determined by a preseted threshold.

\subsubsection{Discussion}
Compared with sparse accumulators, dense accumulator requires a large memory space, especially in high concurrent scenarios. They usually require to maintain a private dense accumulator for each thread or thread group, which greatly reduces the scalability of the algorithm. There are three cases in which dense accumulators are preferred: (1) in hybrid accumulators to process the rows with a large number of \replaced{non-zeros}{non-zero elements} per row; (2) in SpGEMM algorithms on CPU whose large cache and modest computing cores are preferred by dense accumulator; (3) in SpGEMM whose result matrix has small column size. In contrast, the sparse accumulators have lower memory requirement and are more suitable for SpGEMM whose result matrix has a large number of columns. At the same time, sparse accumulators are also the primary choice for GPU-based SpGEMM algorithm, because GPU has limited shared memory. Among sparse accumulators, hash-based accumulators have low memory space requirement, with the cost of handling collisions. For three list-based sparse accumulators, the performance of the used sorting algorithm determines their performance. Heap and merge sort has the same time complexity, while the latter presents a higher space complexity. Radix \replaced{sort}{sorting} is preferred when processing large-scale array.

\section{Architecture Oriented Optimization}\label{sec:Architecture}
Different computer architectures usually have different computing and memory features, so the optimization of SpGEMM is closely related to the architecture used. In the following sections, we introduce SpGEMM optimization for five popular architectures: CPU, GPU, FPGA, ASIC, heterogeneous\replaced{,}{ platform} and distributed platform. Figure \ref{fig:architectures} summarizes existing \replaced{contributions}{researches} on different architectures.

\begin{figure}[htbp]
    \footnotesize
    \centering
    \begin{displaymath}
    \mbox{Architectures}\left\{
    \begin{array}{lr}
    \mbox{CPU-\cite{Patwary2015Parallel}\cite{Chen2019}\cite{gu_bandwidth_2020}\cite{2020_Neural_Sunway}\cite{Elliott2018}}  \\ 
    \mbox{GPU-\cite{zachariadis_accelerating_2020}\cite{Weifeng14}\cite{Liu2015framework}\cite{xia2021scaling}\cite{BellDO12}\cite{Demouth2012}\cite{2020_ICDE_Optimization}\cite{2020_PPoPP_spECK}\cite{Gremse2015} \cite{Deveci2018multilevelmemory}\cite{Liu2019}\cite{DaltonOB15}\cite{Kurt2017} \cite{Nagasaka2017}\cite{Winter2019}\cite{rajamanickam2021kokkoskernels}\cite{TileSpGEMM}} \\ 
    \mbox{FPGA-\cite{5681425}\cite{Jamro2014}\cite{haghi_fp-amg_2020}} \\ 
    \mbox{ASIC-\cite{OuterSPACE2018}\cite{IPDPS2021}\cite{ExTensor2019}\cite{zhang_gamma_2021}\cite{2019TACO_MetaStrider}\cite{2020_MICRO_MatRaptor}\cite{2020_HPCA_SpArch}} \\
    \mbox{Heterogeneous}
        \left\{
            \begin{array}{lr}
                \mbox{CPU+GPU-\cite{Siegel2010}\cite{2019_THPC_Performance}\cite{Matam2012}\cite{Liu2015framework}\cite{Rubensson:2016:LPB:2994680.2994743}} \\
                \mbox{CPU+FPGA-\cite{2020CPUFPGA}} \\
                \mbox{Others-\cite{Sparta2021}} \\
            \end{array}
        \right. \\ 
    \mbox{Distributed-\cite{rasouli2021compressed}\cite{Ballard2013_communication}\cite{hussain2021communication}\cite{Buluc2008}\cite{Buluc2010}\cite{Aydin2011}\cite{Selvitopi2019}\cite{AzadBBDGSTW15}\cite{Jin2004}\cite{Akbudak2014}\cite{Demirci2019}\cite{Weber2015}\cite{Borstnik2014}} \\
    \end{array}
    \right.
    \end{displaymath}
    \caption{A summary of existing researches on different architectures.}
    \label{fig:architectures}
\end{figure}

\subsection{CPU-based Optimization}
\subsubsection{Optimization for Memory Access}
Patwary et al. \cite{Patwary2015Parallel} propose to divide $\matr{A}$ and $\matr{B}$\deleted{ matrices} by row and column, respectively, aiming to alleviate the problem of high L2 cache misses caused by dense accumulator. Elliott et al. \cite{Elliott2018} present a single-phase OpenMP variant of the Gustavson algorithm. They use page-aligned allocations and track used size and capacity separately to accelerate the allocation of thread-local memories. Gu et al. \cite{gu_bandwidth_2020} use the propagation block technique to improve the memory bandwidth utilization of \replaced{OP}{outer-product} SpGEMM. They first save the multiple tuples generated by the outer product into multiple partially sorted bins, and then sort and merge each bin independently. Chen et al. \cite{Chen2019}\cite{2020_Neural_Sunway} design multiple SpGEMM kernels based on different sparse formats. They all utilize coalesced DMA transmission instead of discrete memory access to achieve fast data loading from main memory. Furthermore, they reserve a minimum partition of $\matr{B}$ in local data memory to improve data reuse and avoid redundant data loading.

\subsubsection{Optimization for Load Balance} 
In \cite{Patwary2015Parallel}, matrices are divided into small partitions and dynamic load balancing is used over the partitions. They find that better results are achieved when the total number of partitions is $6$-$10$ times the number of threads. Chen et al. \cite{Chen2019}\cite{2020_Neural_Sunway} present a three-level partitioning scheme for Sunway architecture. At the first level, each core group in the SW26010 processor performs the multiplication of a sub-$\matr{A}$ and $\matr{B}$. At the second level, each CPE core in a core group multiplies the sub-$\matr{A}$ by a sub-$\matr{B}$. At the third level, to fit the 64 KB local data memory in each CPE core, sub-$\matr{A}$ and sub-$\matr{B}$ are further partitioned into several sets. To achieve a balanced load, they divide \replaced{$\matr{A}$ and $\matr{B}$ }{$\matr{A}$ and $\matr{B}$ into sub-$\matr{A}$ and sub-$\matr{B}$, respectively, }according to the computational loads rather than \deleted{equal }rows.

\subsubsection{Optimization for Data Structure}
Patwary et al. \cite{Patwary2015Parallel} store each column partition of $\matr{B}$ in an individual CSR format. This requires to change the data structure of $\matr{B}$ from CSR to blocked CSR in advance, which brings a significant format conversion overhead. Therefore, they use a simple upper-bound method to estimate the proportion of \replaced{non-zeros}{non-zero elements} per row that exceeds the size of the L2 cache, and format conversion occurs only when this proportion is higher than 30\%.

\subsection{GPU-based Optimization}
GPU has emerged as a promising computing device to \replaced{HPC}{high performance computing} for its massive parallelism and high memory bandwidth. On the one hand, a GPU has thousands of streaming processors (SP). Application optimization on GPU should address the problem of scalable parallelism and load unbalancing\deleted{ on thousands of SPs}. On the other hand, GPU has a hierarchical memory architecture including thousands of registers, on-chip shared memories and caches, and local and global memory. Most GPU architecture-oriented optimizations aim to reduce the memory traffic between on-chip and global memory.

\subsubsection{Optimization for Memory Access}
In \cite{DaltonOB15}, Dalton et al. propose an improved ESC method, which stores the \replaced{non-zeros}{non-zero elements} of $\matr{A}$ in shared memory to avoid repeated loading from global memory. In \cite{Winter2019}, Winter et al. propose an adaptive chunk-based GPU SpGEMM approach (AC-SpGEMM), which improves ESC algorithm by reducing intermediate results of \deleted{matrix }$\matr{C}$ within shared memory. Gremse et al. \cite{Gremse2015} also use shared memory to merge intermediate results. The main idea is to reduce the overhead of global memory accesses by merging rows using sub-warps. Then in \cite{Nagasaka2017}, Nagasaka et al. propose a fast SpGEMM algorithm that only requires a small amount of memory and achieves high performance. Since the memory subsystems of NVIDIA$^{\text{\textregistered}}$ GPU differ from generation to generation significantly in both bandwidth and latency, two different data placement methods are proposed in \cite{Deveci2018multilevelmemory}. One is to keep the partitions of $\matr{A}$ and $\matr{C}$ in fast memory and stream the partitions of $\matr{B}$ to shared memory. The other is the opposite. \replaced{Which method to choose}{The method to be chosen} often depends on the features of the input matrices. In \cite{Weifeng14}\cite{Liu2015framework}, memory pre-allocation for the result matrix is organized using a hybrid approach that saves a large amount of global memory space.

\subsubsection{Optimization for Load Balance}
In \cite{Demouth2012}, one warp is assigned to one row of $\matr{A}$, and each thread in the warp is responsible for the multiplication of one non-zero entry in the row and the corresponding row of $\matr{B}$. To address the load imbalance between thread blocks, Lee et al. \cite{2020_ICDE_Optimization} propose \textit{Block Reorganizer} based on the \replaced{OP SpGEMM}{outer-product formulation}. The \textit{Block Reorganizer} first gets the calculation \replaced{load}{workload} of all column-row pairs of input matrices, then combines, splits and assigns these pairs to thread blocks. Kurt et al. \cite{Kurt2017} also use one thread block to process multiple rows of $\matr{A}$ simultaneously. Yu et al. \cite{TileSpGEMM} store input matrices and output matrix as multiple non-empty tiles and assign one warp to process one sparse tile of $\matr{C}$ in both symbolic and numeric phases on GPU, which greatly alleviates the load imbalance caused by irregular non-zero distributions of matrix rows.

To achieve a better load balancing, some work uses thread block to process the same NNZ. In \cite{Winter2019}, Winter et al. split the \replaced{non-zeros}{non-zero elements} of $\matr{A}$ uniformly and assign each thread block to process the same NNZ. However, the workload for each block varies based on the intermediate products generated due to the irregular non-zero distribution in the rows of $\matr{B}$. They present a fine-grained load balancing strategy within thread block. In addition, some researches assign tasks by intermediate results. For example, the authors of \cite{Liu2015framework}\cite{Weifeng14} group the rows of the result matrix to multiple bins according to upper-bound NNZ estimation of each row, and assign different computing units to each bin to have a better load balancing.

In recent years, some authors have suggested two-level load balancing. SpECK \cite{2020_PPoPP_spECK} takes advance of both global load balance which splits the work into blocks, and local load balance which decides the number of threads assigned for each row of $\matr{B}$. Xia et al. \cite{xia2021scaling} also adopt this load balance strategy in their own GPU implementation of SpGEMM. Also, SpGEMM in Kokkos Kernels \cite{rajamanickam2021kokkoskernels} picks the best task assignment method according to non-zeros distribution of input matrices. Specifically, for the matrices with fewer \replaced{non-zeros}{non-zero entries} per row, it assigns one thread to a row. For the matrices with more \replaced{non-zeros}{non-zero entries}, multiple threads are assigned to one row, and vector parallelism is used. In \cite{DaltonOB15}, SpGEMM is formulated as a layered graph model, and the expansion phase of ESC method is considered as the BFS in the levels of the layered mode. Then, they parallel the expansion phase over the multiplication of each non-zero \replaced{entry}{element} of $\matr{A}$ and the corresponding row of $\matr{B}$ at the granularity of \deleted{a }thread, \deleted{a }warp, or \deleted{a }thread block, according to the length of the row referenced from $\matr{B}$.

The emerging tensor core unit (TCU) in GPU attracts the attention of researchers. In \cite{zachariadis_accelerating_2020}, Zachariadis et al. utilize TCUs to improve the performance of SpGEMM. They partition the input matrices into tiles and operate only on tiles which contain one or more \added{non-zero }entries.

\subsubsection{Optimization for Data Structure} Liu et al. \cite{Weifeng14}\cite{Liu2015framework} present efficient parallel merging methods for rows with a large number of intermediate results. It consists of four steps: binary search and duplicate entries reducing, prefix-sum scan, non-duplicate entries copy, and merging of two sorted sequence in one continuous memory space. Dalton et al. \cite{DaltonOB15} propose an optimization scheme for the sorting process of ESC method. They replace global memory-based sorting operations over a large number of intermediate results with multiple parallel shared memory-based sorting operations over a small number of ones. When \replaced{the shared memory space is insufficient}{the number of intermediate entries processed by one thread block exceeds the maximum shared memory space}, the global memory-based sorting is launched. Liu et al. \cite{Liu2019} utilize GPU registers to optimize three typical sparse accumulators: sort, merge, and hash.\deleted{ They perform critical operations (sorting, merging, and hashing) of these accumulators in GPU registers to accelerate the accumulating process.}

\subsection{Field-Programmable Gate Array (FPGA)}
FPGA is an alternative computing device to CPU and GPU, and it has customizable data paths, flexible memory and massive parallel computing units. Lin et al. \cite{5681425} were the first to address SpGEMM on FPGAs. Their design allows user-tunable power-delay and energy-delay trade off by employing different number of processing elements in architecture design and different block size in blocking decomposition.\deleted{ They study the performance of their design in terms of computational latency, as well as the associated power-delay and energy-delay trade off.} The authors of \cite{Jamro2014} believe that the comparisons of the indexes dominate the performance of SpGEMM. Therefore, they propose a highly parallel architecture, which can carry out at least 8$\times$8 indexes comparison in a single clock cycle. Their design uses the \replaced{CbC SpGEMM}{column-by-row formulation}. Moreover, two different formats are used respectively for matrix $\matr{A}$ (CSR) and $\matr{B}$ (CSC). The work of \cite{haghi_fp-amg_2020} proposes an FPGA-based reconfigurable framework FP-AMG that can be reused for all kernels in AMG. The proposed architecture is scalable and reconfigurable, but the implementation details about SpGEMM are not given.

DSP blocks and blocked memory are important components in the design of SpGEMM on FPGA. The number of DSP blocks and the size of blocked memory limit the scale of the dedicated hardware. The problem size is usually very large in practice, and therefore memory blocks are frequently swapped in and out because of small on-chip memories. However, the operating frequency of FPGA generally is lower than that of the CPU or GPU, and the data transmission is also time-consuming. These are the two main reasons why FPGA is not widely employed for SpGEMM.

\subsection{ASIC}
OuterSPACE \cite{OuterSPACE2018} is the first work that addresses the ASIC acceleration of SpGEMM. The authors explore an outer-product based matrix multiplication and revised sparse format in their implementation. Sriseshan et al. \cite{2019TACO_MetaStrider} present a memory-centric architecture MetaStrider to address the fundamental latency-bound inefficiencies of sparse data reduction. SpArch \cite{2020_HPCA_SpArch} is proposed to jointly optimize input and output data, which uses outer-product formulation to reuse input data and \added{uses }on-chip partial matrix merging to reuse output matrix. Different from the outer product, the authors of \cite{2020_MICRO_MatRaptor} present a new architecture (MatRaptor) based on row-wise product. It accesses the sparse data in a vectorized and streaming fashion. The work in \cite{zhang_gamma_2021} presents a SpGEMM accelerator Gamma that also leverages Gustavson algorithm. It uses an on-chip storage structure FiberCache to capture data reuse patterns and supports thousands of concurrent fine granularity fibers. It also achieves high throughput using a dynamic scheduler and preprocessing algorithm. Based on the observation that the most compact memory format and the most efficient computing format need not be the same, the authors of \cite{IPDPS2021} propose an accelerator extension that supports efficient format conversion and optimal format-pair prediction. ExTensor \cite{ExTensor2019} is an approach for accelerating generalized tensor algebra using hierarchical and compositional intersection. There is a metadata engine inside that aggressively looks ahead in the computation to remove useless computations before they are delivered to the arithmetic units. Generally speaking, hardware accelerators prefer algorithms that have small memory footprints. Thereby, row-wise or inner and outer jointly production are more popular than others in current implementations. Additionally, most work reported are evaluated using software simulators, such as GEM5 \cite{OuterSPACE2018}\cite{2020_MICRO_MatRaptor} or home-made cycle-accurate simulators \cite{2020_HPCA_SpArch}\cite{zhang_gamma_2021}\cite{ExTensor2019}\cite{IPDPS2021}. 

It is interesting to see that \replaced{six out of seven papers of ASIC optimization}{most work of ASIC optimization} are reported after \deleted{the year of }2020.\deleted{ There are 16 papers published in 2020 and 10 published in 2021.} However, there is only one paper that presents dedicated hardware design for SpGEMM. The number of paper about SpGEMM has rapidly increased in recently years. We are expecting to see customized chips coming out in the near future.

\subsection{Heterogeneous Platform}

\textbf{CPU+GPU}. Siegel et al. \cite{Siegel2010} develop a task-based programming model and a runtime-supported execution model to achieve dynamic load balancing on CPU-GPU heterogeneous system. For a CPU-GPU heterogeneous computing system, Matam et al. \cite{Matam2012} propose several heuristic methods to identify the proper work division of a subclass of matrices between CPU and GPU.
Liu etl al. \cite{Liu2015framework}\cite{2019_THPC_Performance} exploit heterogeneous processor AMD A10-7850K APU, which shares the system RAM between CPU and GPU, to improve the performance of memory re-allocation when the first size prediction fails for some matrices with relatively long rows. Besides, in \cite{Rubensson:2016:LPB:2994680.2994743}, Rubensson et al. present a method for parallel block SpGEMM on distributed memory clusters based on their previously proposed Chunks and Tasks model \cite{Rubensson2014}. 

\textbf{CPU+FPGA}. The authors of \cite{2020CPUFPGA} introduce a cooperative CPU+FPGA architecture REAP. The sparse matrices are reorganized on CPU and then streamed into FPGA. The CPU performs the symbolic analysis and packs the \replaced{non-zeros}{non-zero entries} in an intermediate representation to increase regularity.

\textbf{Heterogeneous memory architecture}. Liu et al. \cite{Sparta2021} explore the optimization of two sparse tensors contraction (SpTC), a high-order extension of SpGEMM in essence, using the persistent memory-based heterogeneous memory. Based on the knowledge of the SpTC algorithm and data object characteristics, they statically prioritize the data placement between DRAM and persistent memory module (PMM) to achieve the better performance.

\subsection{Distributed Platform} 
Researches of SpGEMM on the distributed\deleted{-memory} platform usually aim at minimizing inter-node communication and simultaneously maintaining balanced load among multiple nodes. In \cite{Jin2004}, Jin et al. propose multiple static and dynamic smart scheduling policies for sparse matrix multiplication under their proposed super programming model (SPM). The distributed SpGEMM implementations provided in the EpetraExt and Tpetra packages of Trilinos \cite{Trilinos-overview} divide the first \deleted{sparse }matrix $\matr{A}$ into multiple computing nodes by row, and then each computing node accesses the needed columns of the second \deleted{sparse }matrix $\matr{B}$ that correspond to the column distribution of the divided $\matr{A}$ rows from a remote location.

\begin{figure}[!htbp]
    \centering
    \includegraphics[width=0.7\textwidth]{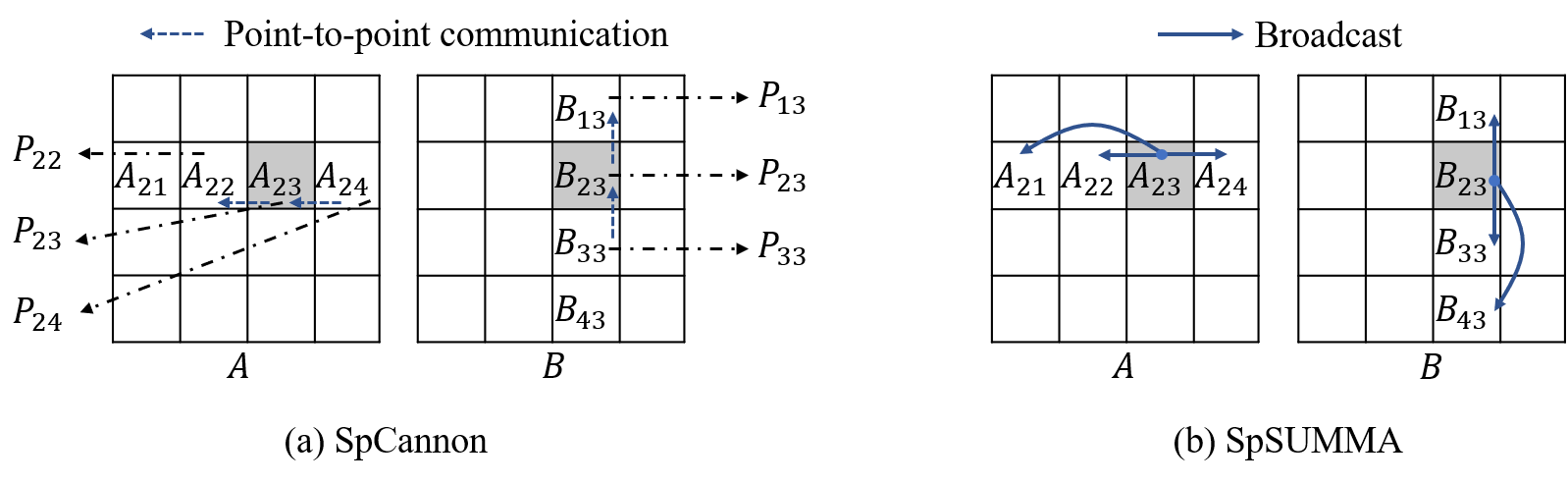}
    \caption{SpCannon and SpSUMMA. Take the processor $P_{23}$ as an example. (a) \textbf{SpCannon}: $P_{23}$ sends $\matr{A}_{23}$ to its left neighbor $P_{22}$ and $\matr{B}_{23}$ to its upper neighbor $P_{13}$, and receives $\matr{A}_{24}$ from its right neighbor $P_{24}$ and $\matr{B}_{33}$ from its lower neighbor $P_{33}$. (b) \textbf{SpSUMMA}: $P_{23}$ broadcasts $\matr{A}_{23}$ along the second row and $\matr{B}_{23}$ along the third column.}
    \label{fig:SpSUMMA}
\end{figure}

To improve the performance of \replaced{distributed SpGEMM}{SpGEMM for distributed computation}, Bulu{\c{c}} et al. \cite{Buluc2008} propose two 2D algorithms: SpSUMMA and SpCannon. For both algorithms, $P$ processors are logically organized on a $\sqrt{P}\times\sqrt{P}$ mesh, and matrices $\matr{A}$, $\matr{B}$, and $\matr{C}$ are assigned to processors according to 2D decomposition. In SpCannon, each processor sends and receives $\sqrt{P}-1$ point-to-point messages of size $nnz(\matr{A})/P$, and $\sqrt{P}-1$ messages of size $nnz(\matr{B})/P$. Instead of the nearest-neighbor communication in SpCannon, row-wise and column-wise broadcasts are used in SpSUMMA. Figure \ref{fig:SpSUMMA} shows examples of SpCannon and SpSUMMA. Later, they extend the above algorithms to the distributed platform with thousands of processors using newer MPI version \cite{Buluc2010}\cite{Aydin2011}. A communication scheme similar to \cite{Aydin2011} is used in \cite{Borstnik2014}, which provides a distributed blocked compressed sparse row (DBCSR) library aiming to accelerate SpGEMM\deleted{ calculations} in the solving of self consistent field (SCF) equations from quantum chemistry.

In the batched 3D SpSUMMA algorithm proposed by Hussain et al. \cite{hussain2021communication}, when the memory requirement to compute the output exceeds available memory of each processor, it accesses each block of $\matr{B}$ batch-by-batch. Split-3D-SpGEMM presented by Azad et al. \cite{AzadBBDGSTW15} splits each two-dimensional sub-matrix into $c$ slices along the third process grid dimension. Rasouli et al. \cite{rasouli2021compressed} propose a new divide-and-conquer SpGEMM. It executes data from previous processor while communicating its data with neighbors to reduce communication time. Ballard et al. \cite{Ballard2013_communication} analyze the lower bounds of bandwidth and latency of distributed 1D/2D/3D algorithms, and propose improved algorithms to lower the communication cost. Weber et al. \cite{Weber2015} organize all processes as a 3D Cartesian topology. Each block $\matr{A}_{ij}$ is located on a specific process according to physical properties of the practical problem. The process that holds $\matr{C}_{ij}$ block receives the blocks $\matr{A}_{ik}$ and $\matr{B}_{kj}$ from other processes, and performs $\matr{C}_{ij}=\matr{C}_{ij}+\matr{A}_{ik}\matr{B}_{kj}$.

Selvitopi et al. \cite{Selvitopi2019} use two-constraint hypergraph and bipartite graph models to enable balancing processors' loads in both map and reduce phases, and minimizing data transfer in shuffle phase. Demirci et al. \cite{Demirci2019} design a distributed SpGEMM algorithm on Accumulo. It alleviates multiple times scanning of the input matrices by using Accumulo's batch scanning capability. They also propose a bipartite graph-based partition scheme to reduce the total communication volume and provide a balance of workload among servers. In \cite{Akbudak2014}, Akbudak et al. propose a two-phases parallel SpGEMM algorithm, which uses a two-constraint hypergraph partitioning to guide partitioning for maintaining a balanced load over two phases. In the first phase, each processor owns a column stripe of $\matr{A}$ and a row stripe of $\matr{B}$, and then finishes the communication-free local SpGEMM computations. The second phase reduces partial results yielded in the first phase to calculate the final value.

\subsection{Discussion
}
In view of the importance of SpGEMM in classical scientific computing and its wide application in graph analysis, ASIC-based SpGEMM optimization has been increasing in recent years. CPU-based SpGEMM optimization work mainly focuses on reasonable partition of the input matrices to ensure high cache hits and load balancing among computing units. The SpGEMM optimization on GPU usually uses registers and shared memory to speedup multiplication and accumulation as much as possible, so as to reduce the access to global memory. In view of multi-level parallelism (thread, warp, and thread block), assigning different workloads to different parallel levels to ensure load balancing has also attracted great attention of researchers. On heterogeneous platforms, an efficient partitioning method that can fully utilize the computing power of both CPU and accelerator is one of the important designing objectives. Balanced workload and minimal communication overhead among computing processes are critical on distributed platforms.

\section{Programming model}\label{sec:progmmingModel}
A programming model is a bridge between a programmer's logical view and physical view of program execution on some specific hardware \cite{BerkeleyView}. Over the years, a number of programming models were proposed and only a few of them are widely used. Most of these popular programming models are either driven by commercial companies or embraced by open-source communities. Table \ref{tab:program_model} lists popular programming models and frameworks reported in the literature. 

\begin{table}[h]
\centering
\caption{A summary of different programming models.}
\label{tab:program_model}
\scalebox{0.75}{
\begin{tabular}{ll|ll} 
\toprule
 \textbf{Model} & \makebox[6cm][c]{\textbf{Contribution}} & \textbf{Model} & \makebox[6cm][c]{\textbf{Contribution}}  \\
 \midrule
 Tasks-based & \cite{Siegel2010}\cite{Rubensson2014}\cite{Jin2004}\cite{NAGASAKA2019102545}\cite{Rubensson:2016:LPB:2994680.2994743}\cite{Yasar2018} & OpenCL & \cite{Matam2012}\\
 \midrule
 OpenMP & \cite{rasouli2021compressed}\cite{hussain2021communication}\cite{Matam2012}\cite{NAGASAKA2019102545}\cite{Elliott2018}\cite{2020_TPDS_Cartesian} &MPI & \cite{rasouli2021compressed}\cite{Ballard2013_communication}\cite{Borstnik2014}\cite{Akbudak2018}\cite{Chen2019}\cite{Elliott2018}\cite{2020_Neural_Sunway}\cite{Buluc2010}\cite{Aydin2011}\cite{rajamanickam2021kokkoskernels}\\
 \midrule
 CUDA &\cite{Gremse2015}\cite{Liu2019}\cite{Matam2012}\cite{2019_THPC_Performance}\cite{2020_ICDE_Optimization}\cite{2020_PPoPP_spECK}\cite{BellDO12}\cite{Demouth2012}\cite{Nagasaka2017}\cite{xia2021scaling}\cite{TileSpGEMM} & MapReduce & \cite{Selvitopi2019} \\
\bottomrule
\end{tabular}}
\end{table}

\subsection{Task-based programming model}
Task-based programming model is a high-level abstraction in parallel computing. In this model, \replaced{a workload}{the computation work} is partitioned into smaller tasks recursively until a certain task granularity is reached. The computation of the program follows a fork-join model that concurrent tasks fork at designated points and join later at a subsequent point. Each task is executed on one processing unit.

There are several task-based programming models used in SpGEMM implementations, and they have different paradigms. Cilk is a revised-C/C++ language for parallel computing. It uses two keywords \textit{spawn} and \textit{sync} to orchestrate tasks in the computation, and uses a work-stealing algorithm to balance task in the runtime. Paper \cite{Yasar2018} proposes KKTri-Clik algorithm and uses a heuristic strategy to find a balanced partition. Instead of a language extension like Cilk, Threading Building Blocks (TBB) \cite{intelTBB2008} explores a library-based implementation. The library manages task mapping and scheduling at runtime. It is only used for memory allocation/deallocation to gain higher performance on multi-core and many-core processors in \cite{NAGASAKA2019102545}. Paper \cite{Siegel2010} defines a task-based programming framework that supports partitioning the SpGEMM in blocks to address the problem of load balancing. \deleted{The authors of \cite{Rubensson2014} also define a new task-based programming model. }Different from \cite{Siegel2010}, the authors of \cite{Rubensson2014} propose Chunks and Tasks, a new task-based programming model. It maps the chunks and tasks to physical resources. Paper \cite{Rubensson:2016:LPB:2994680.2994743} also uses this programming model in their implementation. There are also some other customized task libraries for SpGEMM \cite{Jin2004}. However, these libraries are only used in the reported work.

\subsection{OpenMP and MPI}
OpenMP is one of the most popular programming paradigms to enable threaded parallelism \cite{openmp}. It uses \deleted{a number of }preprocessing directives to tell compilers which code block can be executed in parallel. OpenMP is much easier to use than the task-based model. However, programmers have to make sure that their programs using OpenMP are data race free. The authors of \cite{NAGASAKA2019102545} target Intel Xeon Phi architecture and use OpenMP to parallelize loops. In the domain of high performance computing, OpenMP is usually used together with MPI, which is a specification for distributed computing API that enables many computers to communicate with and work together. Generally, SpGEMM on distributed systems endorses a two-level parallelism that MPI facilitates communication among SMP nodes and OpenMP manages multiple-threads on each SMP node \cite{rasouli2021compressed}\cite{hussain2021communication}\cite{Elliott2018}\cite{Borstnik2014}\cite{2020_TPDS_Cartesian}. Almost all the work that reported on super computers (Cray XT4 \cite{Aydin2011}, BlueGene/Q system \cite{Akbudak2018}, Sunway TaihuLight \cite{Chen2019}\cite{2020_Neural_Sunway}, Astra and Fugaku \cite{rajamanickam2021kokkoskernels}) and clusters \cite{Buluc2010} use MPI.

OpenMP is also used for CPU+X heterogeneous architectures, in which X can be any hardware accelerators, such as GPU \cite{Matam2012} and KNL. In general, both host and device are used for computation and heuristics are designed to find the balanced work division between CPU and \replaced{X}{GPU}. 

\subsection{MapReduce}
MapReduce is a structured parallel programming model proposed by Google that serves for processing large data sets in a massively parallel manner \cite{MapReduce}. It is a very important programming pattern that is supported in the Hadoop framework based on the Hadoop file system. Paper \cite{Selvitopi2019} has its SpGEMM implementation atop MR-MPI, an open-source implementation of MapReduce written for distributed\deleted{-memory parallel} machines on top of standard MPI. \replaced{This work schedules}{The motivation of this work is to schedule} map and reduce tasks statically in a MapReduce job to \replaced{improve}{achieve} data locality and load balance.

\subsection{CUDA and OpenCL}
Both CUDA and OpenCL can be used to program GPU devices to maximize data parallelism using the SIMT programming model. Their difference lies in that CUDA is from NVIDIA$^{\text{\textregistered}}$ and OpenCL is an open standard supporting many devices from different vendors, such as CPU, GPU and DSP. The most used versions in existing work include CUDA 4.0 \cite{BellDO12}\cite{Matam2012}, CUDA 6.0\cite{Gremse2015}, CUDA 8.0 \cite{Liu2019}\cite{xia2021scaling} and CUDA 10.x \cite{2020_PPoPP_spECK}. CUDA uses multiple streams to express concurrency and stream sequence of operations to GPU devices. Some reported work uses multiple streams to overlap not only executions, but also data transfers \cite{Matam2012}. Results reported in \cite{xia2021scaling} confirm the effectiveness of launching multiple CUDA kernels with CUDA streams for each group to execute concurrently.

CUDA and OpenCL facilitate the programming of massively parallel computing devices. Nevertheless, a deep understanding of the underlying architecture is essential to have high performance gain. Especially after the new Volta architecture is released, enabling independent threads scheduling, and it's more challenging to write correct code on new GPU devices. 

In addition, some libraries also provide the Python wrappers that bridge the gap between C/C++ and Python. For example, Pygraphblas \cite{pygraphblas} makes it easy and simple to call GraphBLAS \cite{graphblas} APIs in Python. PyTrilinos \cite{pytrilinos} allows Python developers to import Trilinos \cite{trilinos-website} packages and then call their APIs in a Python program.

\section{Performance Evaluation}\label{sec:Evaluation}

\begin{table}[htp]
        \centering
        \caption{Detailed information of evaluated libraries.}
        \scalebox{0.75}{
        \begin{tabular}{lllclll}
            \toprule
            \textbf{Device} & \textbf{Library} & \textbf{Version} & \textbf{Open-source} & \textbf{Size prediction} &\textbf{Accumulator} & \textbf{Format}\\
            \midrule 
            \multirow{2}{*}{CPU} & KK-OpenMP\cite{rajamanickam2021kokkoskernels} & 3.5.00 & \cmark & precise & dense & CSR  \\ \cmidrule(lr){2-7}
            & MKL\cite{MKL}   & 2020.4.304 & \xmark & precise & - & CSC/CSR/BSR \\
            \midrule
            \multirow{7}{*}{GPU} & cuSPARSE\cite{cuSPARSE}  & 11.4.120 & \xmark & precise & hash & CSR\\
            \cmidrule(lr){2-7}
            & CUSP\cite{CUSP} & 0.5.0 & \cmark & upper-bound  & list & COO \\ \cmidrule(lr){2-7}
            & bhSPARSE\cite{Liu2015framework}\cite{Weifeng14}     & 2015.11.6 & \cmark & hybrid  & hybrid & CSR \\ \cmidrule(lr){2-7}
            & KK-CUDA\cite{rajamanickam2021kokkoskernels} & 3.5.00 & \cmark & precise & hash & CSR  \\ \cmidrule(lr){2-7}
            & NSparse\cite{Nagasaka2017} & 1.5.0 & \cmark & precise & hash & CSR \\ \cmidrule(lr){2-7}
            & spECK\cite{2020_PPoPP_spECK} & 2022.1.2 & \cmark & precise & hybrid & CSR \\ \cmidrule(lr){2-7    }
            & TileSpGEMM\cite{TileSpGEMM} & 2022.1.25 & \cmark & precise & hybrid & Tiled structure  \\
            \midrule
            \multirow{3}{*}{\makecell[c]{Distributed\\ system}} &  CTF\cite{CTF-Sparse} & 1.5.5 & \cmark & precise & dense & CSR \\
            \cmidrule(lr){2-7}
            & PETSc & 3.17.3 & \cmark & precise & list & CSR \\
            \bottomrule 
        \end{tabular} }
        \label{tab:library}
\end{table}
\subsection{Overview}
In this section, we conduct a series of experiments to compare the SpGEMM performance of several state-of-art implementations on three platforms, including CPU, GPU, and distributed system. Table \ref{tab:library} lists details of the libraries evaluated in each part. CPU and GPU-based SpGEMM implementations in Kokkos Kernels are labeled with KK-OpenMP and KK-CUDA, respectively.

\subsection{System Setup and Benchmark}
CPU-based SpGEMM is tested on a machine running 64-bit Ubuntu 18.04 and equipped with one Intel$^{\text{\textregistered}}$ Xeon$^\text{\textregistered}$ E5-2680 v4 with 2.40 GHz clock frequency and 14 physical cores, supporting 28 threads. For GPU-based libraries, two different NVIDIA$^{\text{\textregistered}}$ GPUs are used. The first is Tesla P100, which is based on Pascal architecture and equipped with 3,584 CUDA cores and 16 GB device memory. The second is Tesla V100, which is based on the Volta architecture and equipped with 5,120 CUDA cores and 16 GB device memory. The version of CUDA Toolkit is 11.4. \replaced{The performance evaluation for distributed SpGEMM is conducted}{We conduct the performance evaluation for two distributed SpGEMM implementations} on a cluster including 10 nodes, each of which is equipped with two Intel$^{\text{\textregistered}}$ Xeon$^\text{\textregistered}$ Gold 6258R with 2.7 GHz clock frequency and 56 physical cores\deleted{ in total}.

\replaced{All the tested sparse matrices}{The sparse matrices used in our tests} are downloaded from the SuiteSparse Matrix Collection \cite{Davis2011}. We use the same dataset for performance tests on CPU and GPU. It has 1,880 square and 678 rectangular \deleted{sparse }matrices. Considering the powerful computing power and non-negligible communication overhead of distributed system, we choose 10 \deleted{sparse }matrices whose \replaced{NNZ}{number of non-zero elements} is greater than 5M\deleted{ from the SuiteSparse Matrix Collection}. Table \ref{tab:distri-dataset} lists details of these matrices. SpGEMM benchmark of $\matr{A}\times \matr{A}$, commonly used in Markov clustering algorithm\deleted{ (please refer to Algorithm \ref{alg:MarkovClustering})}, is used for all square matrices. The benchmark of $\matr{A}\times \matr{A}^T$, universal in AMG solver, is used for all rectangular matrices. The performance of both \textit{single} and \textit{double} floating-point is tested.

\begin{table}[!ht]
    \centering
    \caption{Tested sparse matrices for distributed system.}
    \scalebox{0.75}{
    \begin{tabular}{lccc|lccc}
    \toprule
        \textbf{Matrix} & \textbf{M} & \textbf{N} & \textbf{NNZ} & \textbf{Matrix} & \textbf{M} & \textbf{N} & \textbf{NNZ} \\ \midrule
        TSOPF\_RS\_b2052\_c1 & 25,626 & 25,626 & 6,761,100  & Chebyshev4 & 68,121 & 68,121 & 5,377,761 \\ \midrule
        TSOPF\_RS\_b678\_c2 & 35,696 & 35,696 & 8,781,949  & test1 & 392,908 & 392,908 & 12,968,200\\ \midrule
        largebasis & 440,020 & 440,020 & 5,560,100  & PR02R & 161,070 & 161,070 & 8,185,136\\ \midrule
        marine1 & 400,320 & 400,320 & 6,226,538  & ohne2 & 181,343 & 181,343 & 11,063,545\\ \midrule
        Goodwin\_127 & 178,437 & 178,437 & 5,778,545  & torso1 & 116,158 & 116,158 & 8,516,500\\ \bottomrule
    \end{tabular}}
    \label{tab:distri-dataset}
\end{table}

\subsection{Evaluation Results on CPU}

\begin{table}[h]
\centering
\caption{Comparison of the number of sparse matrices for which each algorithm achieves the best performance (the left side of "/") and runs successfully (the right side of "/") on CPU.}
\label{tab:numberMatrices_CPU}
\scalebox{0.75}{
\begin{tabular}{c|rrr|rr} 
    \toprule
        \multirow{2}{*}{\textbf{Precision}} & \multicolumn{3}{|c|}{\textbf{\# of matrices}} & \multicolumn{2}{c}{\textbf{Speedup to KK-OpenMP}} \\ \cmidrule(lr){2-6}
        & MKL-CSR & MKL-BSR & KK-OpenMP & MKL-CSR & MKL-BSR \\
    \midrule
    single & 1,187/2,342 & 993/1,235 & 216/2,396 & 3.46 & 13.64  \\ \midrule
    double & 1,256/2,304 & 921/1,212 & 219/2,396 & 3.56 & 15.34  \\
    \bottomrule
\end{tabular}}
\end{table}

\begin{figure}[htbp]
	\centering
	\subfloat[Single] 
	{	\includegraphics[width=0.44\linewidth]{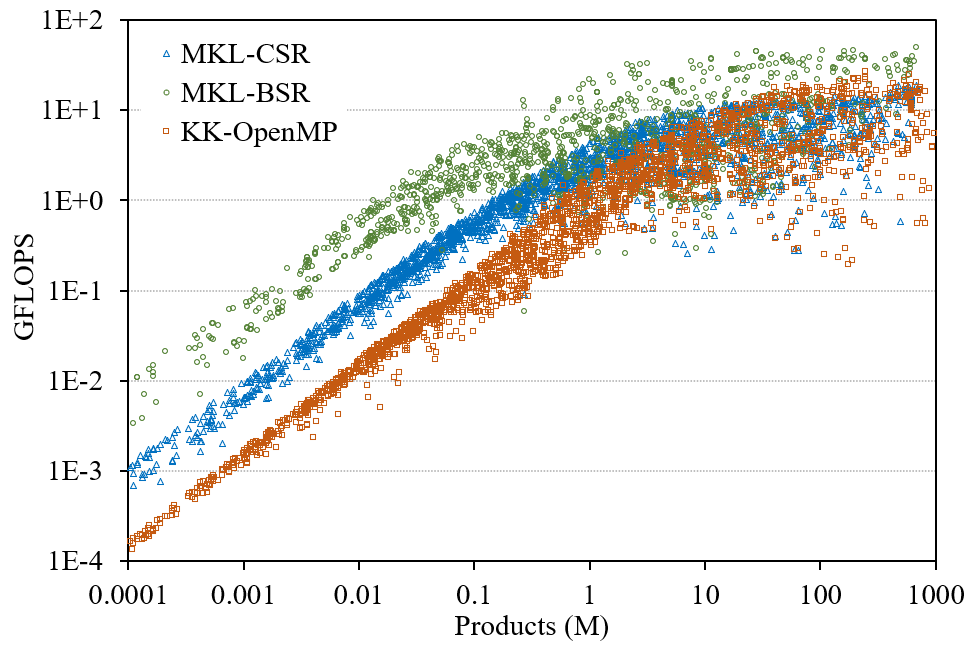}
	}
	\subfloat[Double] {
		\includegraphics[width=0.44\linewidth]{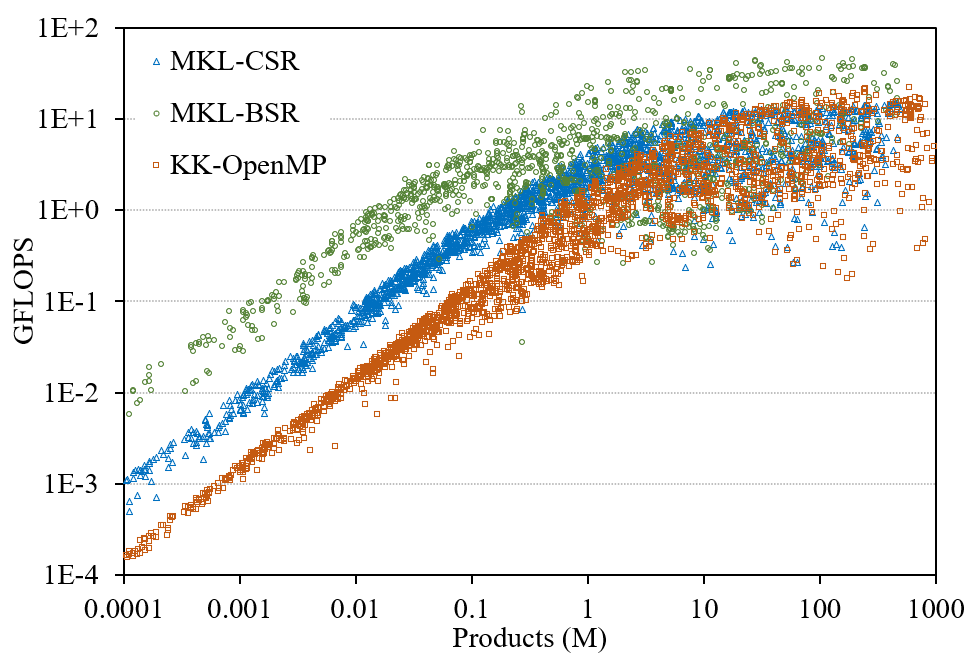}
	}
	\caption{GFLOPS on CPU ordered by the total number of products.}
	\label{fig:CPU-GFLOPS}
\end{figure}

MKL uses a highly encapsulated internal data structure to perform operations on sparse matrices. We encode both input matrices in either CSR or BSR format and call \textit{mkl\_sparse\_sp2m} to perform SpGEMM. The MKL implementation supports two-stage execution. The row pointer array of the output matrix is calculated in the first stage. In the second stage, the remaining column indexes and value arrays of the output matrix are calculated. We consider the execution time of these two stages as the run time of SpGEMM in MKL, while format conversion between CSR/BSR and internal representation is considered as the preprocessing overhead and discussed later. Besides, we set 28 threads for KK-OpenMP. Figure \ref{fig:CPU-GFLOPS} presents the performance of three SpGEMM algorithms on CPU ordered by the number of products, which equals to the upper-bound predicted NNZ. Table \ref{tab:numberMatrices_CPU} lists the number of sparse matrices for which each algorithm presents the best performance and runs successfully on CPU, as well as the average speedup of MKL-CSR and MKL-BSR to KK-OpenMP. We make the following observations: 
\begin{itemize}[listparindent=-0.5cm,leftmargin=0.3cm,topsep=0.1cm]
    \item MKL-BSR and MKL-CSR show better performance than KK-OpenMP for most sparse matrices, and achieve an average speedup of more than 3x and 13x for both precision, respectively. Specifically, KK-OpenMP achieves the highest GFLOPS for about 8\% matrices in single and double precision.
    \item MKL-BSR \deleted{outperforms MKL-CSR for a considerate number of sparse matrices and }achieves the highest GFLOPS for more than 900 sparse matrices in both precision, but it fails to run on about 53\% matrices. The reason is that the format conversion from default CSR to BSR poses a large memory requirement.
    \item KK-OpenMP presents a comparable performance to MKL-CSR and MKL-BSR for large-scale matrices. The reason is that small-scale matrices can not fully utilize the powerful hardware, and their performance gain from multi-threaded parallel computing are offset by thread creation and synchronization. On the contrary, MKL-CSR runs successfully for more than 90\% matrices, and shows the best performance for about half of them.
\end{itemize}

We also evaluate the performance of single \deleted{floating-point }precision on CPU. For MKL-CSR, the average and maximum speedups of single to double precision are 1.05x and 2.48x, respectively. MKL-BSR \deleted{shows better performance in single precision and }achieves an average speedup of 1.17x on average. In addition, an average speedup of 1.09x and a maximum speedup of 2.33x are achieved for KK-OpenMP.

\subsection{Evaluation Results on GPU} 
\begin{figure}[htbp]
	\centering
	\subfloat[Single] 
	{
		\includegraphics[width=0.44\linewidth]{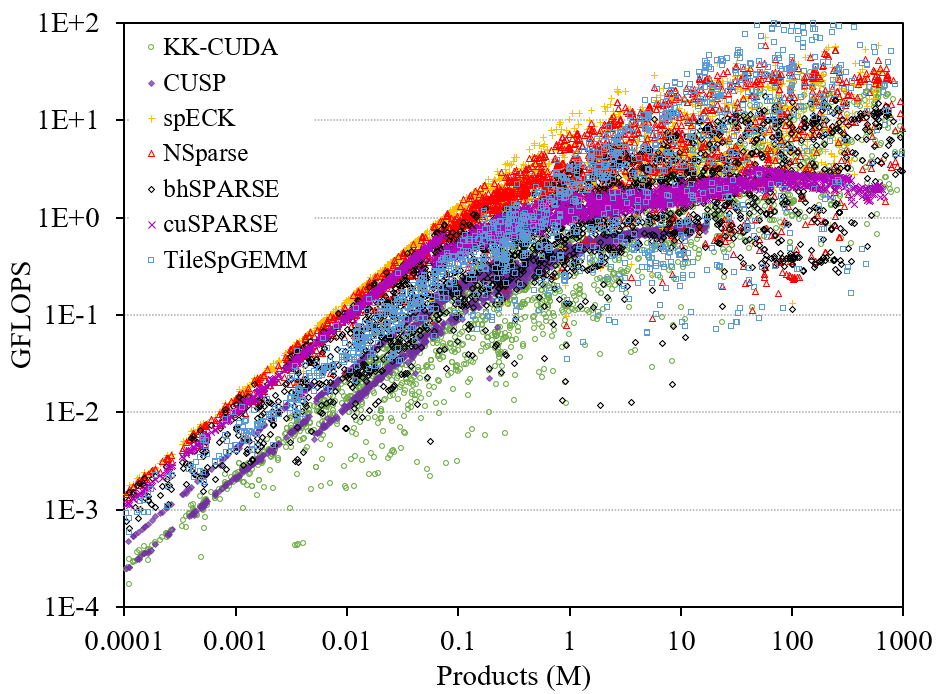}
	}
	\subfloat[Double] {
		\includegraphics[width=0.44\linewidth]{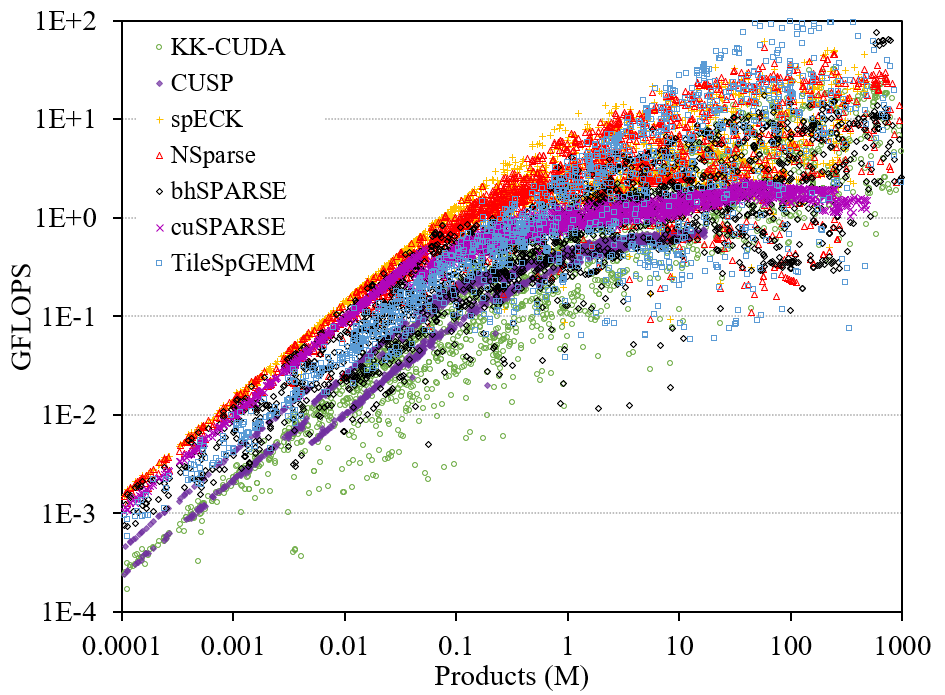}
	}
	\caption{GFLOPS on Tesla P100 ordered by the total number of products.}
	\label{fig:GPU-GFLOPS-P100}
\end{figure}

\begin{figure}[htbp]
	\centering
	\subfloat[Single] 
	{
		\includegraphics[width=0.44\linewidth]{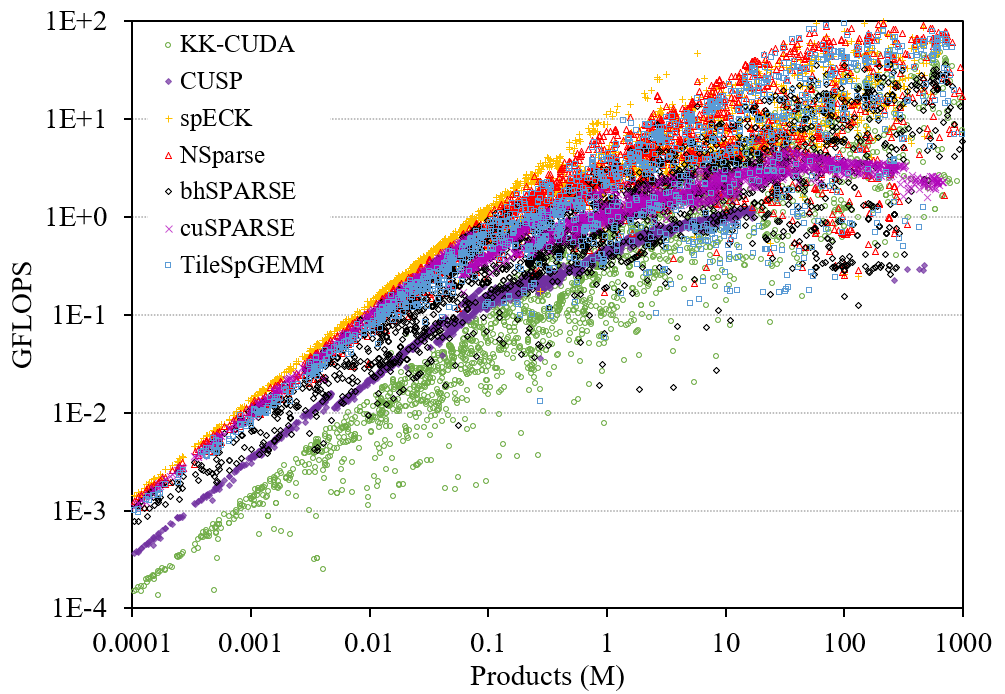}
	}
	\subfloat[Double] {
		\includegraphics[width=0.44\linewidth]{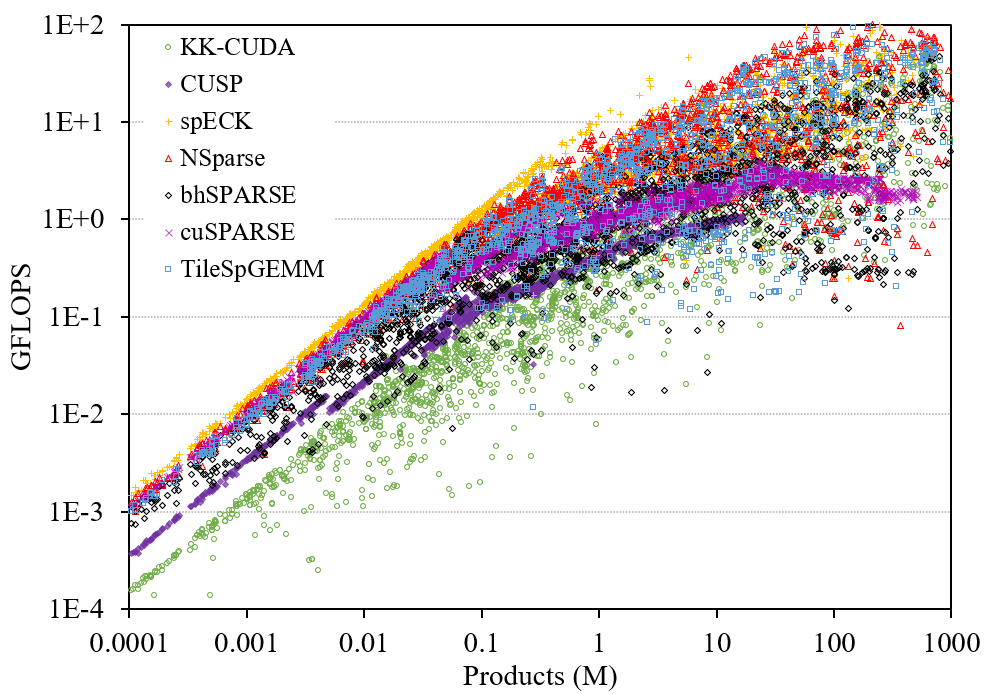}
	}
	\caption{GFLOPS on Tesla V100 ordered by the total number of products.}
	\label{fig:GPU-GFLOPS-V100}
\end{figure}

\begin{table}[htbp]
\centering
\caption{Comparison of the number of sparse matrices for which each algorithm achieves the best performance (the left side of "/") and runs successfully (the right side of "/") on Tesla P100 and V100.}
\label{tab:numberMatrices_GPU}
\scalebox{0.7}{
\begin{tabular}{ccrrrrrrr} 
    \toprule
     \multirow{2}{*}{\textbf{GPU}} & \multirow{2}{*}{\textbf{Precision}} & \multicolumn{7}{c}{\textbf{SpGEMM algorithm}} \\ \cmidrule(lr){3-9}
     & & CUSP & cuSPARSE & NSparse & spECK & bhSPARSE & KK-CUDA & TileSpGEMM \\
     \midrule
    \multirow{2}{*}{P100} & single & 0/1,848 & 43/2,347 & 791/2,352 & 1,201/2,333 & 0/2,375 & 16/2,391 & 343/1,689 \\ \cmidrule(lr){2-9}
    & double & 0/1,845 & 33/2,301 & 798/2,317 & 1,165/2,281 & 12/2,383 & 28/2,388 & 357/1,689 \\ \midrule
    \multirow{2}{*}{V100} & single & 0/1,853 & 29/2,346 & 837/2,367 & 1,337/2,335 & 2/2,373 & 19/2,391 & 170/1,689 \\ \cmidrule(lr){2-9}
    & double & 0/1,848 & 17/2,305 & 847/2,378 & 1,338/2,335 & 5/2,385 & 18/2,391 & 170/1,695 \\
    \bottomrule
\end{tabular}}
\end{table}

\begin{table}[htbp]
\centering
\caption{Average speedup to CUSP on Tesla P100 and V100.}
\label{tab:speedup_GPU}
\scalebox{0.7}{
\begin{tabular}{ccrrrrrrr} 
    \toprule
     \multirow{2}{*}{\textbf{GPU}} & \multirow{2}{*}{\textbf{Precision}} & \multicolumn{6}{c}{\textbf{SpGEMM algorithm}} \\ \cmidrule(lr){3-8}
     & & cuSPARSE & NSparse & spECK & bhSPARSE & KK-CUDA & TileSpGEMM \\
     \midrule
    \multirow{2}{*}{P100} & single & 3.70 & 7.06 & 7.79 & 2.69 & 1.25 & 4.36 \\ \cmidrule(lr){2-8}
    & double & 3.49 & 7.33 & 8.02 & 2.77 & 1.40 & 4.85 \\ \midrule
    \multirow{2}{*}{V100} & single & 3.25 & 6.56 & 7.78 & 2.66  & 0.86 & 5.02 \\ \cmidrule(lr){2-8}
    & double & 2.96 & 6.56 & 7.55 & 2.49 & 0.88 & 4.81 \\
    \bottomrule
\end{tabular}}
\end{table}

\begin{table}[h]
\centering
\caption{Average speedup of single to double precision on Tesla P100 and V100.}
\label{tab:precision_GPU}
\scalebox{0.7}{
\begin{tabular}{c|rrrrrrr} 
    \toprule
     \multirow{2}{*}{\textbf{GPU}} & \multicolumn{7}{|c}{\textbf{SpGEMM algorithm}} \\ \cmidrule(lr){2-8}
     & CUSP & cuSPARSE & NSparse & spECK & bhSPARSE & KK-CUDA & TileSpGEMM \\
     \midrule
    P100 & 1.09 & 1.23 & 1.08 & 1.11 & 1.08 & 1.00 & 1.03 \\ \midrule
    V100 & 1.05 & 1.20 & 1.06 & 1.08 & 1.13  & 1.02 & 1.08 \\
    \bottomrule
\end{tabular}}
\end{table}

Seven GPU-based SpGEMM algorithms are evaluated in this section. Figures \ref{fig:GPU-GFLOPS-P100} and \ref{fig:GPU-GFLOPS-V100} present their performance on Tesla P100 and V100, respectively. Table \ref{tab:numberMatrices_GPU} lists the number of matrices for which each algorithm achieves the best performance and runs successfully. Moreover, the average speedup to CUSP is given in Table \ref{tab:speedup_GPU}.  We summarize our observations in the following points:
\begin{itemize}[listparindent=-0.5cm,leftmargin=0.3cm,topsep=0.1cm]
    \item CUSP and TileSpGEMM run successfully for about 72\% and 68\% matrices on both GPUs. This is because the SpGEMM implementation of CUSP uses the ESC method, which requires a large memory space to store the results of scalar multiplication and to support the sorting operation. TileSpGEMM uses a tile format, which requires the input matrix to be square. Therefore, it can not process all rectangular sparse matrices, which account for about 27\% of the entire dataset.
    \item On both GPUs, CUSP does not present the best performance for all matrices, while spECK and NSparse achieve the best performance for most sparse matrices. TileSpGEMM outperforms other SpGEMM algorithms for 10\%$\sim$20\% successfully run matrices. In addition, cuSPARSE and KK-CUDA achieve the highest GFLOPS for similar number of matrices on both GPUs, and bhSPARSE is superior to other SpGEMM algorithms for a few matrices on both GPUs.
    \item KK-CUDA shows different performance on two GPUs compared with other SpGEMM algorithms. Specifically, its performance on Tesla P100 is better than that of CUSP, and it achieves an average speedup to CUSP of 1.25x and 1.40x in single and double precision, respectively. On Tesla V100, however, its overall performance is inferior to that of all other SpGEMM algorithms.
\end{itemize}

We also compare the performance of each algorithm with different floating-point precision, and Table \ref{tab:precision_GPU} lists the average speedup of single to double precision SpGEMM. We can observe that the single-precision SpGEMM runs slightly faster than double-precision SpGEMM on both two GPUs. Specifically, the average speedups of all SpGEMM algorithms fall within the interval [1.00, 1.23] on Tesla P100 and [1.02,1.20] on Tesla V100. Although single-precision SpGEMM runs faster than double-precision SpGEMM on most sparse matrices, the difference is not significant, especially for KK-CUDA, its performance of single precision is very close to that of double precision on both GPUs. We find that for some SpGEMM algorithms such as spECK and KK-CUDA, the important parameters are determined according to the floating point precision used. It means that the performance comparison between two floating-point precision also includes the impact of parameters change. On the other hand, the idea of two-stage calculation is used in most SpGEMM algorithms. The first stage calculates the size of the result sparse matrix, which does not involve any floating-point calculation. When the symbolic phase accounts for a relatively large proportion of the overall execution time, the change in floating point precision has less impact on performance.

\begin{figure}[!htbp]
	\centering
	\subfloat[Single] 
	{
		\includegraphics[width=0.47\linewidth]{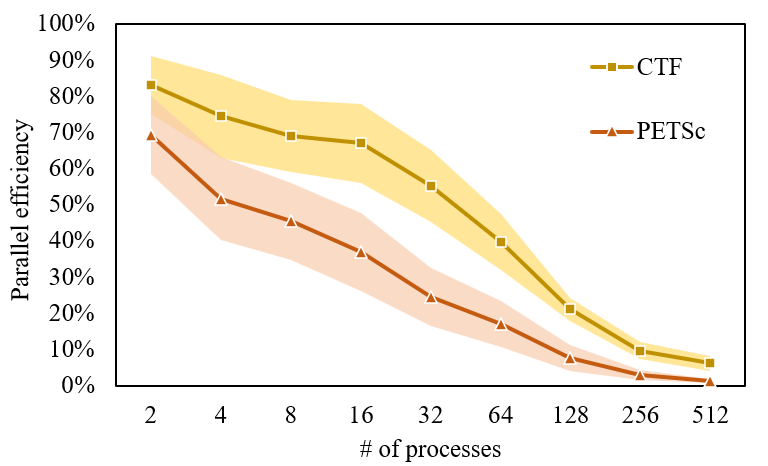}
	}
	\subfloat[Double] {
		\includegraphics[width=0.47\linewidth]{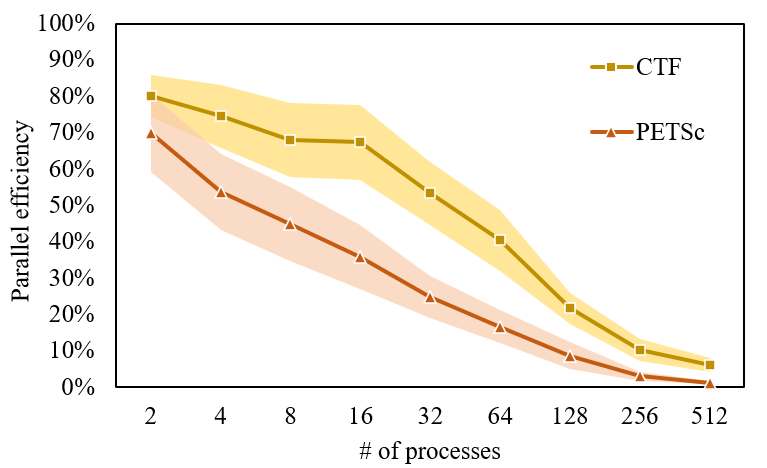}
	}
	\caption{Parallel efficiency on distributed system. Each line is the average parallel efficiency over all sparse matrices displayed with a 95\% colored confidence interval.}
	\label{fig:distri-PE}
\end{figure}

\begin{table}[htbp]
    \centering
    \caption{Average speedup over 1-process PETSc on distributed system.}
    \scalebox{0.7}{
    \begin{tabular}{c|rrrrrrrrr|rrrrrrrrrr}
    \toprule
        \textbf{SpGEMM} & \multicolumn{9}{|c|}{\textbf{PETSc}} & \multicolumn{10}{|c}{\textbf{CTF}} \\ \midrule
        Processes & 2 & 4 & 8 & 16 & 32 & 64 & 128 & 256 & 512 & 1 & 2 & 4 & 8 & 16 & 32 & 64 & 128 & 256 & 512 \\ \midrule
        single & 1.39  & 2.07  & 3.64  & 5.91  & 7.83  & 10.91  & 9.81  & 7.52  & 7.02  & 0.29  & 0.40  & 0.65  & 1.27  & 2.37  & 4.65  & 7.37  & 7.11  & 4.90  & 5.36  \\ \midrule
        double & 1.40  & 2.15  & 3.59  & 5.74  & 7.98  & 10.69  & 11.06  & 7.79  & 6.72  & 0.34  & 0.52  & 0.91  & 1.59  & 3.14  & 6.00  & 10.41  & 10.88  & 7.50  & 7.72 \\ \bottomrule
    \end{tabular}}
    \label{tab:distri_speedup}
\end{table}

\subsection{Evaluation Results on Distributed System}
The performance of CTF and PETSc on distributed system is evaluated in this section. Figure \ref{fig:distri-PE} summarizes their parallel efficiency, and their run time is compared in Table \ref{tab:distri_speedup}, which lists the average speedup to 1-process PETSc. It can be observed that CTF achieves higher parallel efficiency than PETSc in almost all processes settings, but its average performance is inferior to that of PETSc. For single precision, the average speedup of PETSc with all processes settings is higher than that of CTF. For double precision, however, the average speedup of CTF is close to that of PETSc at 64, 128, and 256 processes, and higher \deleted{than that of PETSc }at 512 processes. Moreover, we find that CTF achieves a higher maximum speedup than PETSc for single-precision SpGEMM when the number of processes is larger than 32. For double-precision SpGEMM, it holds for all tested settings. In summary, PETSc achieves the better average performance than CTF, but CTF shows a better parallel efficiency and is expected to outperform PETSc when a large number of processes are available. Besides, we also compare the performance of CTF and PETSc with different floating-point precision. The experimental results from CTF show that the average speedup of single to double over all processes settings falls within the interval [1.04,1.11]. The interval becomes [1.21,1.40] for PETSc.

\subsection{Evaluation Results of Preprocessing Overhead}

Since other SpGEMM algorithms do not has a preprocessing stage, we only present the preprocessing overhead of MKL-CSR, MKL-BSR, and TileSpGEMM in this section. MKL-CSR includes two preprocessing operations: storing an input CSR matrix using the internal data structure (referred to as \textit{import}), exporting the output matrix from the internal structure to the CSR \deleted{format }(referred to as \textit{export}). Although MKL-BSR \replaced{is superior to}{achieves higher performance than} MKL-CSR on a considerable number of matrices, it also requires extra preprocessing operation: format converting between CSR and BSR (referred to as \textit{convert}). In TileSpGEMM, two input sparse matrices encoded with the CSR \deleted{format }are required to be compressed using the tile structure (\replaced{referred to}{abbreviated} as \textit{CSR2Tile}), and the output matrix compressed with the tile structure is also required to be converted to the CSR \deleted{format }(\replaced{referred to}{abbreviated} as \textit{Tile2CSR}).

Experimental results show that, for both MKL-CSR and MKL-BSR, the overhead of \textit{export} is negligible, and \textit{import} takes less than one SpGEMM on average. The \textit{convert} in MKL-BSR is more time-consuming than other operations, taking about 8 SpGEMM on average. In TileSpGEMM, the format conversion between the tile structure and the CSR requires careful consideration, especially \textit{CSR2Tile}, because it costs tens of single TileSpGEMM for some matrices. \textit{Tile2CSR} has less time overhead than \textit{CSR2Tile}, about 5 SpGEMM on average.

From the above discussion, we can conclude that most of the preprocessing overhead of SpGEMM comes from the \deleted{compression }format conversion. MKL-BSR, TileSpGEMM, and CUSP use BSR, tile structure, and COO format, respectively. All other SpGEMM algorithms use the popular CSR\deleted{ format}. Our evaluation on preprocessing overhead shows that the proposal of a new format requires careful consideration. It is necessary to evaluate the performance of SpGEMM in specific applications, simultaneously considering the overhead of format conversion.

\section{Challenges and Future Work}\label{sec:Challenge}
SpGEMM has gained a lot of attention in recent decades, and the work has been conducted in many directions. We believe that it will be used in many other fields as the development of IoT, big data, and AI. We summarize some potential research directions and challenges as follows.

\textbf{Optimization based on machine learning\added{ (ML)}}. One of the challenges of SpGEMM is exploring the sparsity of sparse matrices. Researchers find that the \replaced{non-zeros}{non-zero entries}' distribution dominates the performance of SpMV and SpGEMM on the same architecture. Over the past years, the parameter auto-tuning and automatic selection of sparse formats and SpMV algorithms based on \replaced{ML}{machine learning} models have been designed for SpMV optimization, and significant performance improvement has been observed. However, \replaced{ML}{machine learning}-based SpGEMM optimization is more challenging due to the sparsity consideration of \replaced{three}{two} input sparse matrices, complicated matrix partitioning, and load balancing. 

\textbf{Optimization of size prediction}. Among the four size prediction methods, upper-bound prediction may result in memory over-allocation, and progressive and probabilistic prediction may lead to memory re-allocation. Both over-\deleted{allocation} and re-allocation are \replaced{time-consuming}{challenging tasks} for acceleration devices, such as GPU, DSP, FPGA, and TPU. On one hand, on-device memory capacity is limited and may not accommodate such a large amount of data. On the other head, it takes time to copy data to and from device memory because of separated memory space. Precise prediction is the most popular method\deleted{ in existing research}. It not only reduces the overhead of memory management, but also indirectly improves the performance of result accumulating. However, we compared the run time of each stage in two-stage SpGEMM algorithms and found that the symbolic stage is even more expensive than the numerical stage for some matrices. Therefore, more efficient size prediction is required.

\textbf{Heterogeneous architecture-oriented optimization}. CPU is good at processing complicated logic, while the GPU is good at dense computations. Besides, DSP and FPGA \deleted{accelerators }may be used in different systems. One of the critical questions of porting SpGEMM to the heterogeneous system is how to achieve load balance and minimize communication traffic. Moreover, \replaced{sub-matrices}{each sub-matrix} may have \replaced{different}{its feature} non-zeros' distribution. Ideally, relatively dense blocks should be mapped to acceleration devices, while ultra sparse blocks can be assigned to CPU. Only in this way can each device give full play to its architecture and optimize the overall computing performance of SpGEMM.

\textbf{Application specific optimization.} Application specific optimization of SpGEMM tends to be more useful and effective. Taking unsmoothed aggregation-based AMG for example, the sparse matrix $\matr{P}$ of Galerkin product $\matr{P}^T\matr{AP}$ is a binary matrix with at most one non-zero \replaced{entry}{element} per row. Therefore, the library AmgX \cite{AMGX_2015} develops a custom kernel to speed up \replaced{the product}{this calculation}. However, in classical AMG, $\matr{P}$ is a tall-skinny sparse matrix. The calculating order of the Galerkin product, $(\matr{P}^T\matr{A})\matr{P}$ or $\matr{P}^T(\matr{AP})$, is important to the performance of hash accumulator-based SpGEMM. Specifically, $\matr{P}^T(\matr{AP})$ tends to have less intermediate results accumulating overhead per row than $(\matr{P}^T\matr{A})\matr{P}$ \cite{AMGX_2015}. Although the SpGEMM optimization based on application characteristics may reduce its extensibility, it is worthwhile if significant performance gains can be achieved. We expect more application characteristics to be fully exploited and utilized in the future.

SpMM is well-supported by existing hardware accelerators and GPU devices, as it is an important kernel of many convolutional algorithms. However, the sparsity of the matrices in convolution is much lower than that in HPC\deleted{ computing}. Moreover, most high-performance computing units (such as tensor cores) only support the calculation of low floating-point precision (e.g. 16-bit half precision) because the precision is not so important in AI applications. This is not true for scientific and engineering computing, in which double precision is usually used for the convergence of solvers. Converting a sparse matrix to a dense one and running it on tensor cores are not promising as there are too much wasting work and extra overhead. SpGEMM is useful in the convolution algorithm if \added{both }the input \replaced{and model are}{is sparse and the model is also} sparse. As the development of AI techniques, model pruning and optimization techniques, SpGEMM is expected to be supported in \replaced{HPC}{high-performance computing} devices.

\section{Conclusion}\label{sec:conclusion}
The design of an efficient SpGEMM algorithm, as well as its implementation, is critical to many large-scale scientific applications. Therefore, it is not surprising that SpGEMM has attracted much attention from researchers over the years. In this survey, we highlight some developments in recent years and emphasize the applications, formulations, challenging problems, architecture-oriented optimizations, programming models, and performance evaluation. Some interesting conclusions can be summarized. Row-by-row is the most commonly used\deleted{ SpGEMM} formulation because of its high parallelism and low cache requirement. CSR is the most frequently used storage format\deleted{ in SpGEMM algorithms}, as it is widely used in various fields and thus avoids expensive format conversion. SpGEMM in MKL presents the best performance on CPU. On GPU, spECK and NSparse outperform others and achieve the best performance on a large number of matrices. TileSpGEMM also shows excellent performance for regular and square \deleted{sparse }matrices\deleted{ on GPU}. On distributed system, PETSc is the winner when the number of processes is small. However, CTF presents better performance as the number of processes increases.

In conclusion, we stress the fact that despite recent progress, there are still important areas where much work remains to be done, such as different formats supporting, application and architecture-oriented optimization, and efficient size prediction\deleted{ of target matrix}. On the other hand, the heterogeneous architecture may give rise to new optimization opportunities, and efficient implementation in the specific architecture is within reach and can be expected to continue in the future.

\begin{acks}
The authors would like to thank Zhaonian Tan and Yueyan Zhao for their early work. We would also like to extend our thanks to all reviewers for their constructive comments and suggestions. This work is supported by the \grantsponsor{GS501100001809}{National Natural Science Foundation of China}{https://doi.org/10.13039/501100001809} under Grant No. \grantnum{GS501100001809}{61972033}.
\end{acks}

\bibliographystyle{ACM-Reference-Format}
\bibliography{ref-CSUR.bib}

\end{document}